%% file: main.tex
\documentclass[12pt,a4paper]{scrartcl}

\pdfoutput=1

\usepackage[utf8]{inputenc}
\usepackage[T1]{fontenc}
\usepackage{amsmath,amssymb,mathtools}
\usepackage[separate-uncertainty=true,retain-unity-mantissa=false]{siunitx}
\usepackage{slashed}
\usepackage{graphicx}
\usepackage{booktabs,multicol,multirow}
\usepackage[shortlabels]{enumitem}
\usepackage{hyperref}
\usepackage[capitalize]{cleveref}
\usepackage{xspace}
\usepackage[noblocks]{authblk}
\usepackage[dvipsnames]{xcolor}
\usepackage{soul}
\usepackage[textsize=tiny]{todonotes}
\usepackage{subcaption}
\usepackage{ulem}
\usepackage{bbold}
\usepackage[sorting=none,%
  citestyle=numeric-comp,%
  bibstyle=mynumeric,%
  giveninits=true]{biblatex}
\input{abbr.tex}

\addtokomafont{disposition}{\rmfamily}

\title{\Large 
Theoretical concepts and measurement prospects\\ for BSM trilinear couplings:\\ a case study for scalar top quarks}
\author[1]{Henning Bahl\footnote{\href{mailto:hbahl@uchicago.edu}{hbahl@uchicago.edu}}}
\author[2]{Johannes Braathen\footnote{\href{mailto:johannes.braathen@desy.de}{johannes.braathen@desy.de}}}
\author[2,3]{Georg Weiglein\footnote{\href{mailto:georg.weiglein@desy.de}{georg.weiglein@desy.de}}}
\affil[1]{University of Chicago, Department of Physics, 5720 South Ellis Avenue, Chicago, IL~60637~USA}
\affil[2]{Deutsches Elektronen-Synchrotron DESY, Notkestr.~85, 22607 Hamburg, Germany}
\affil[3]{II. Institut für Theoretische Physik, Universität Hamburg, Luruper Chaussee 149, 22761 Hamburg, Germany}
\date{}
\titlehead{\hfill \parbox{3cm}{\vspace{-2cm}\begin{flushright} \texttt{DESY-22-208} \end{flushright}}}

\begin{document}
\maketitle

\begin{abstract}\noindent
    \input{sec_abstract.tex}
\end{abstract}
\setcounter{footnote}{0}

\newpage

\tableofcontents

\newpage


\section{Introduction}
\label{sec:intro}

\input{sec_intro.tex}


\section{The stop mixing parameter}
\label{sec:Xt_stop_sector}

\input{sec_Xt_stop_sector.tex}


\section{Measurement of the stop mixing parameter}
\label{sec:Xt_measurement}

\input{sec_Xt_measurement.tex}


\section{Renormalisation of the stop sector}
\label{sec:Xt_renormalization}

\input{sec_Xt_renormalization.tex}


\section{Renormalisation of the stop mixing parameter and the prediction for the mass of the SM-like Higgs boson}
\label{sec:Xt_Mh}

\input{sec_Xt_Mh.tex}


\section{Conclusions}
\label{sec:conclusions}

\input{sec_conclusions.tex}


\section*{Acknowledgements}
\sloppy{We thank Ivan Sobolev for collaboration in the early stages of this work, as well as Pietro Slavich for interesting discussions and helpful comments on our manuscript. 
J.B.\ and G.W.\ acknowledge support by the Deutsche Forschungsgemeinschaft (DFG, German Research Foundation) under Germany's Excellence Strategy -- EXC 2121 ``Quantum Universe'' – 390833306. H.B.\ acknowledges support by the Alexander von Humboldt foundation. This work has been partially funded by the Deutsche Forschungsgemeinschaft (DFG, German Research Foundation) - 491245950.
}


\appendix


\section{Bottom-Yukawa corrections to the \texorpdfstring{$X_t$}{Xt} conversion}
\label{app:bottom}

\input{app_bottom.tex}


\clearpage
\printbibliography

\end{document}

%% file: abbr.tex
\newcommand{\deltaOL}{\delta^{(1)}}

\renewcommand{\Re}{\text{Re}}

\newcommand{\cp}{\ensuremath{{\cal CP}}}

\newcommand{\OS}{{\ensuremath{\text{OS}}}\xspace}
\newcommand{\MS}{{\ensuremath{\overline{\text{MS}}}}\xspace}
\newcommand{\DR}{{\ensuremath{\overline{\text{DR}}}}\xspace}
\newcommand{\MDR}{{\ensuremath{\overline{\text{MDR}}}}\xspace}
\newcommand{\fin}{{\ensuremath{\text{fin}}}\xspace}

\newcommand{\gev}{\;\text{GeV}\xspace}
\newcommand{\tev}{\;\text{TeV}\xspace}

\newcommand{\SUSY}{{\text{SUSY}}}
\newcommand{\MSSM}{{\text{MSSM}}}

\newcommand{\msusy}{\ensuremath{M_\SUSY}\xspace}

\newcommand{\order}[1]{\ensuremath{{\cal O}(#1)}}
\newcommand{\al}{\alpha}
\newcommand{\als}{\al_s}
\newcommand{\alt}{\al_t}
\newcommand{\alb}{\al_b}

\newcommand{\mtL}{m_{\tilde{t}_L}}
\newcommand{\mtR}{m_{\tilde{t}_R}}
\newcommand{\mbL}{m_{\tilde{b}_L}}
\newcommand{\mbR}{m_{\tilde{b}_R}}

\newcommand{\pXt}{\phi_{X_t}}
\newcommand{\xt}{\ensuremath{\widehat X_t}\xspace}
\newcommand{\xb}{\ensuremath{\widehat X_b}\xspace}

\NewBibliographyString{refname}
\NewBibliographyString{refsname}
\DefineBibliographyStrings{english}{%
  refname = {Ref\adddot},
  refsname = {Refs\adddot}
}
\DeclareCiteCommand{\ccite}
  {%
  \ifnum\thecitetotal=1
    \bibstring{refname}%
  \else%
    \bibstring{refsname}%
  \fi%
  \addnbspace\bibopenbracket%
  \usebibmacro{cite:init}%
   \usebibmacro{prenote}}
  {\usebibmacro{citeindex}%
   \usebibmacro{cite:comp}}
  {}
  {\usebibmacro{cite:dump}%
   \usebibmacro{postnote}%
   \bibclosebracket}
\newrobustcmd*{\Ccite}{\bibsentence\ccite}

%% file: sec_abstract.tex
After the possible discovery of new heavy particles at the LHC, it will be crucial to determine the properties and the underlying physics of the new states. In this work, we focus on scalar trilinear couplings, employing as an example the case of the trilinear coupling of scalar partners of the top quark to the Higgs boson. We discuss possible strategies for experimentally determining the scalar top (stop) trilinear coupling parameter, which controls the stop--stop--Higgs interaction, and we demonstrate the impact of different renormalisation prescriptions for this parameter. We find that the best prospects for determining the stop trilinear coupling arise from its quantum effects entering the model prediction for the mass of the SM-like Higgs boson in comparison to the measured value. We point out that the prediction for the Higgs-boson mass has a high sensitivity to the stop trilinear coupling even for heavy masses of the non-standard particles. Regarding the renormalisation of the stop trilinear coupling we identify a renormalisation scheme that is preferred in view of the present level of accuracy and we clarify the source of potentially large logarithms that cannot be resummed with standard renormalisation group methods.

%% file: sec_intro.tex
So far, only one scalar particle without a known substructure has been found: the Higgs boson with a mass of about 125~GeV discovered at the Large Hadron Collider (LHC) in 2012~\cite{ATLAS:2012yve,CMS:2012qbp}. Within the Standard Model (SM) of particle physics, the detected Higgs boson is identified with the Higgs boson that is predicted as the only fundamental scalar in this model. However, in many extensions of the SM by physics beyond the SM (BSM) additional scalar degrees of freedom are introduced in order to address questions that are unresolved in the SM, for instance the nature of dark matter, the origin of neutrino masses, or the observed baryon asymmetry of the universe.

Additional spin-zero particles can either be introduced by extending the SM Higgs sector, for example singlet extensions of the SM, two-Higgs-doublet models, etc., or by adding a completely new scalar sector, for example in supersymmetric (SUSY) theories, which associate a scalar degree of freedom with each fermion degree of freedom. A particular new type of interaction potentially arising in these models is an interaction between three scalars that is not generated by a vacuum expectation value. This type of interaction is forbidden in the SM due to the $SU(2)_L$ gauge symmetry of the model. Correspondingly, the trilinear interaction of the SM Higgs boson is generated only after electroweak symmetry breaking. In BSM theories, trilinear scalar couplings can, however, arise as a consequence of dimensionful couplings independently of spontaneous symmetry breaking.

These dimensionful couplings appear for instance in extensions of the SM Higgs sector by one or more scalar gauge singlet(s) if no $\mathbb{Z}_2$ symmetry is imposed~\cite{Chen:2014ask,Kanemura:2015fra,Bahl:2021rts}. Another example are supersymmetric theories which predict trilinear couplings between the Higgs bosons and the supersymmetric partners of the SM fermions. Among these couplings, the trilinear coupling between the supersymmetric partners of the top quark (which are usually called scalar top quarks or stops) and the SM-like Higgs boson is of particular importance. This ``stop mixing parameter'' is typically the largest among the trilinear couplings and controls not only the Higgs--stop--stop interaction itself but also the mass splitting between the stops.

If an extended scalar sector is discovered at the LHC or a future collider, the measurement of the interactions between the various scalars will be crucial to pinpoint the underlying theory. With this motivation in mind, we discuss in this paper how trilinear scalar couplings should be properly defined in the theoretical predictions and how they can be extracted from experimental measurements. We focus our discussion on the example of the stop mixing parameter of the Minimal Supersymmetric SM (MSSM), building upon earlier work in the literature~\cite{Bartl:1997yi,Berggren:1999ss,Bartl:2000kw,Finch:2002qy,Rolbiecki:2009hk,Desch:2004cu,LHCLCStudyGroup:2004iyd,Djouadi:2013lra,ElKosseifi:2022rkf}. We will also point out aspects of our discussion that are valid for other theories with trilinear scalar couplings. 

We review different ways to extract the stop mixing parameters from experimental measurements. Going beyond existing results, we point out the difficulties of the various approaches and emphasise the crucial role of the mass of the SM-like Higgs boson. In connection with this discussion, we compare different known schemes~\cite{Eberl:1996np,Djouadi:1996wt,Beenakker:1996de,Djouadi:1998sq,Bartl:1997pb,Bartl:1998xp,Guasch:1998as,Kraml:1999qd,Brignole:2001jy,Hollik:2003jj,Heinemeyer:2007aq,Baro:2009gn,Fritzsche:2011nr,Fritzsche:2013fta, Hollik:2014wea, Hollik:2014bua, Passehr:2014xwu} for the renormalisation of the stop mixing parameter in Higgs boson mass calculations, and we examine what scheme choice would be most appropriate. In this context, we ascertain the origin of large Sudakov-like logarithms plaguing the Higgs boson mass calculation in the on-shell scheme when combining diagrammatic and EFT techniques. Based on this discussion, we propose to use a mixed scheme where the stop mixing parameter is renormalised in the \DR/\MDR scheme while the stop masses are renormalised on-shell.

This work is structured as follows. We present a short review of the MSSM stop sector in \cref{sec:Xt_stop_sector}. In \cref{sec:Xt_measurement}, we point out difficulties in measuring the stop mixing parameter in various approaches. Based on this discussion, we review different possibilities to renormalise the stop sector in \cref{sec:Xt_renormalization}. In \cref{sec:Xt_Mh}, we discuss the origin of Sudakov-like logarithms affecting Higgs mass calculations in the on-shell scheme incorporating renormalisation-group resummations. Our conclusions can be found in \cref{sec:conclusions}. \cref{app:bottom} provides additional details regarding the conversion of the stop mixing parameter between the on-shell and the \DR scheme.

%% file: sec_Xt_stop_sector.tex
In the MSSM, trilinear scalar couplings 
can arise from terms in the superpotential --- in the form of the $\mu$ parameter --- as well as from the soft SUSY-breaking Lagrangian --- in the form of the trilinear couplings $A_f$. We focus here on the interaction of stops with Higgs bosons.

As a first consequence of the Higgs--stop--stop interaction, mixing between the superpartners of the left- and right-handed components of the top quark, which we denote by $\tilde t_L$ and $\tilde t_R$, respectively, is induced. This is directly visible in the stop mass matrix which takes the following form,
\begin{equation}
  \label{eq:stop_mass_matrix}
  \mathbf{M}_{\tilde{t}} = 
  \begin{pmatrix}
    m_{\tilde{t}_L}^2 + m_{t}^2 + \cos(2\beta)(\frac{1}{2} - \frac{2}{3}s_W^2) M_Z^2 & m_{t} \; X_{t}^* \\
    m_{t} \; X_{t}   &  m_{\tilde{t}_R}^2 + m_{t}^2 + \frac{2}{3}\cos(2\beta) s_W^2 M_Z^2
  \end{pmatrix},
\end{equation}
where $m_{\tilde{t}_{L,R}}^2$ are the stop soft SUSY-breaking masses, and $X_t \equiv A_t - \mu^*/\tan\beta$ is the stop mixing parameter  ($\tan\beta$ denotes the ratio of the Higgs vacuum expectation values, $\tan\beta \equiv t_\beta \equiv v_u/v_d$). 
$M_Z$ is the mass of the $Z$ boson and $s_W$ the sine of the weak mixing angle. $m_t$ is the top-quark mass, for which we will use a value of $173.2\gev$ throughout this paper.

The stop mass matrix can be diagonalised by a unitary transformation,
\begin{align}\label{eq:stop_LR_12_transformation}
\begin{pmatrix} \tilde t_1 \\ \tilde t_2 \end{pmatrix} = \mathbf{U}_{\tilde t} \begin{pmatrix} \tilde t_L \\ \tilde t_R \end{pmatrix}
\end{align}
such that
\begin{align}
\mathbf{U}_{\tilde t}\mathbf{M}_{\tilde t}\mathbf{U}_{\tilde t}^\dagger = \text{diag}(m_{\tilde t_1}^2, m_{\tilde t_2}^2),
\hspace{.5cm} \text{with}
\hspace{.5cm} \mathbf{U}_{\tilde t} = 
\begin{pmatrix} c_{\tilde t} & s_{\tilde t} e^{-i\phi_{X_t}} \\ - s_{\tilde t} e^{i\phi_{X_t}} & c_{\tilde t} \end{pmatrix},
\hspace{.5cm} \mathbf{U}_{\tilde t}\mathbf{U}_{\tilde t}^\dagger = \mathbb{1}.
\end{align}
Here $m_{\tilde t_1}^2 \leq m_{\tilde t_2}^2$ by definition,
\begin{align}
\phi_{X_t} = \arg(X_t) ,
\end{align}
and we introduced the abbreviations $s_\gamma \equiv \sin\gamma$ and $c_\gamma \equiv \cos\gamma$ for a generic angle~$\gamma$.

The mixing angle $\theta_{\tilde t}$ obeys the relation
\begin{align}
\cos(2\theta_{\tilde t}) = \frac{m_{\tilde t_R}^2 - m_{\tilde t_L}^2 - M_Z^2 c_{2\beta}(\frac{1}{2} - \frac{4}{3}s_W^2)}{m_{\tilde{t}_2}^2 - m_{\tilde{t}_1}^2} ,
\end{align}
where the stop masses are given by
\begin{align}
m_{\tilde t_{1,2}}^2 = m_t^2 + \frac{1}{2}\left\{m_{\tilde{t}_L}^2 + m_{\tilde{t}_R}^2 \mp \sqrt{\left[m_{\tilde{t}_L}^2 - m_{\tilde{t}_R}^2 + M_Z^2 c_{2\beta} \left(\frac{1}{2} - \frac{4}{3}s_W^2\right)\right]^2 + 4 m_t^2 |X_t|^2}\right\}.
\end{align}
For later use, we define the abbreviations
\begin{align}
    \msusy = \sqrt{m_{\tilde t_L} m_{\tilde t_R}}, \hspace{.5cm} \widehat X_t = X_t/\msusy\,.
\end{align}
The Higgs--stop--stop interaction is, however, not only manifest in the stop mass matrix but also induces a direct coupling of the light \cp-even Higgs boson $h$ to two stops. In the limit of vanishing electroweak gauge couplings, these couplings read
\begin{align}
c(h \tilde t_1 \tilde t_1) &= - 2 i \frac{m_t}{v} (m_t + s_{\theta_{\tilde t}} c_{\theta_{\tilde t}} |X_t|), \label{eq:hSt1St1_coupling}\\
c(h \tilde t_1 \tilde t_2) &= - i \frac{m_t}{v} c_{2\theta_{\tilde t}} |X_t| e^{i\phi_{X_t}}, \label{eq:hSt1St2_coupling}\\
c(h \tilde t_2 \tilde t_2) &= - 2 i \frac{m_t}{v} (m_t - s_{\theta_{\tilde t}} c_{\theta_{\tilde t}}|X_t|), \label{eq:hSt2St2_coupling}
\end{align}
where $v \simeq 246\gev$ denotes the vacuum expectation value (vev), with $v^2 = v_u^2 + v_d^2$.

%% file: sec_Xt_measurement.tex
As discussed in \cref{sec:Xt_stop_sector}, the stop mixing parameter $X_t$ induces mixing between the left- and right-handed stops. Therefore, observables depending on the stop mixing are also sensitive to the stop mixing parameter. Since the stop mixing parameter $X_t$ also appears directly in the Higgs--stop--stop interaction, measuring processes involving a Higgs boson and two stops as external particles allows one to directly constrain $X_t$ without resorting to its relation to the stop mixing angle. Moreover, the Higgs--stop--stop interaction can additionally influence other observables at the quantum level, such as the predictions for the Higgs boson masses. We note that the observations in this section (with the exception of the Higgs mass predictions) also apply to other BSM theories with trilinear couplings between three different particles.

In the following, we will discuss these different possibilities in detail and qualitatively assess their prospects for experimentally determining the parameter $X_t$ at the LHC and future high-energy colliders. We will in the following mainly restrict ourselves to the case where the parameter $X_t$ is real. Prospects for measuring the phase of $X_t$ are discussed in \ccite{Rolbiecki:2009hk}.


\subsection{Stop masses}

\begin{figure}
\centering
\begin{minipage}{.48\textwidth}
\includegraphics[width=\textwidth]{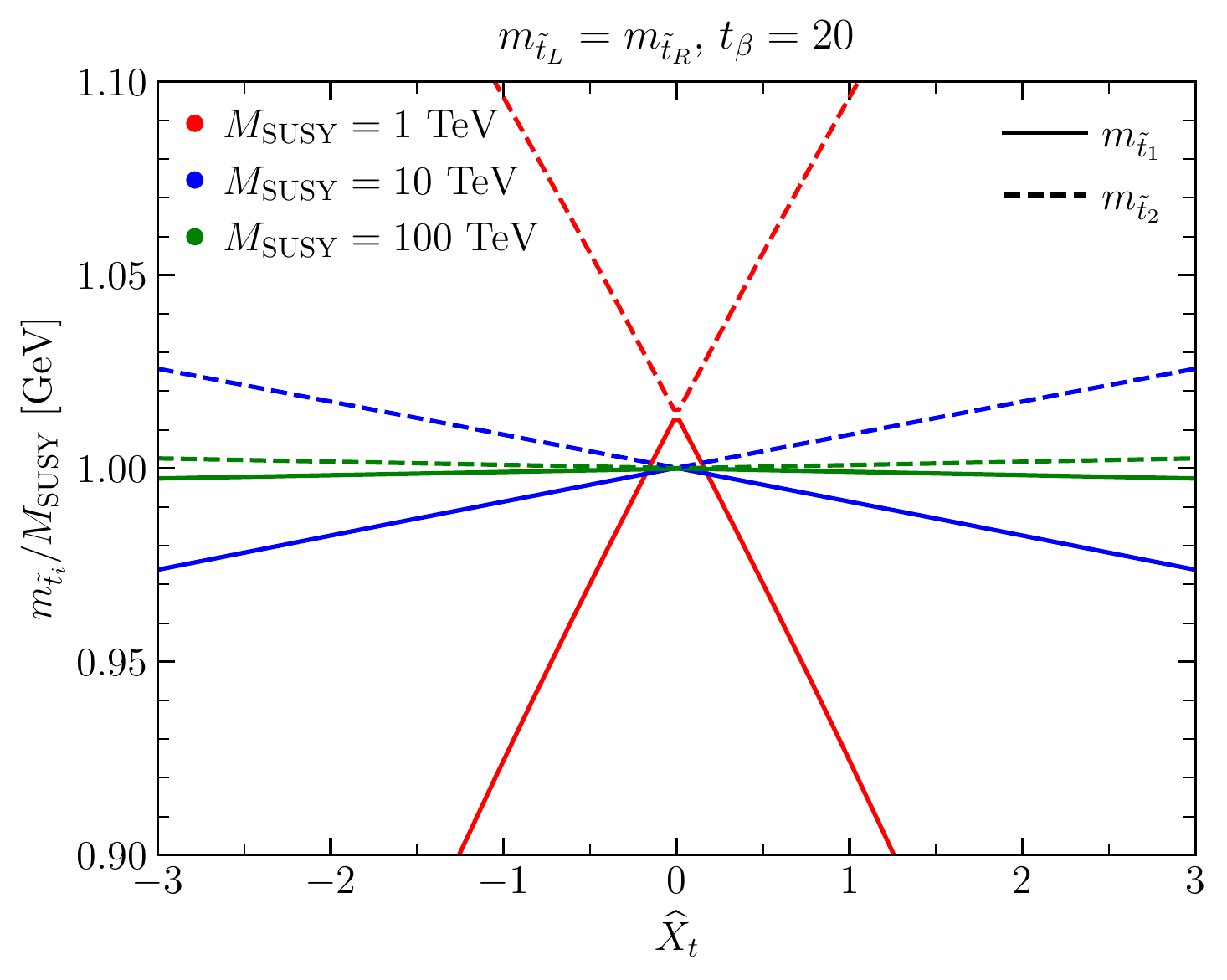}
\end{minipage}
\begin{minipage}{.48\textwidth}
\includegraphics[width=\textwidth]{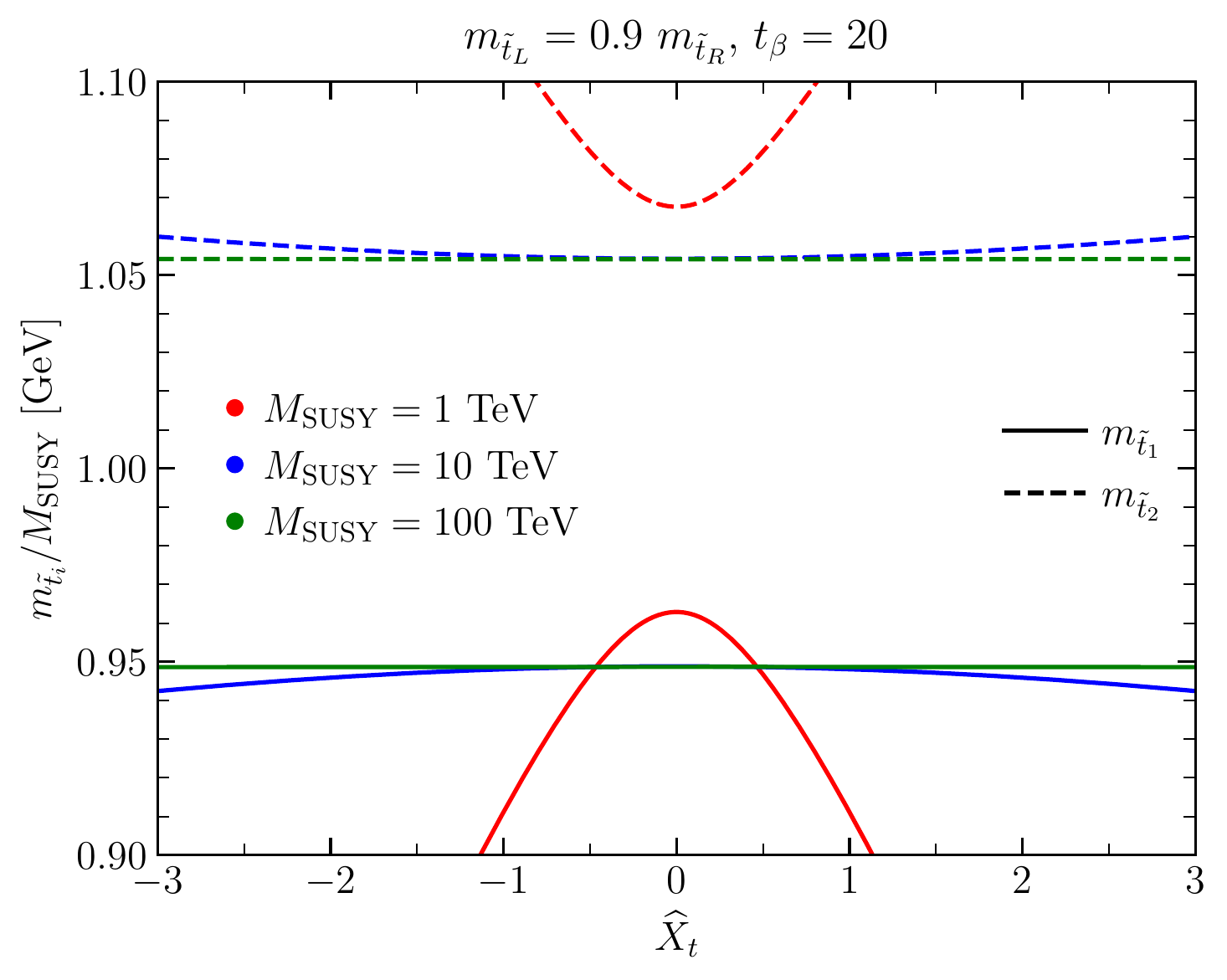}
\end{minipage}
\caption{Dependence of the (tree-level) stop masses on $\widehat X_t$ for $m_{\tilde{t}_L} = m_{\tilde{t}_R}$ (left) and $m_{\tilde{t}_L} = 0.9\, m_{\tilde{t}_R}$ (right) for $\tan\beta = 20$. Results for the light (solid curves) and heavy (dashed curves) stop mass eigenstates, normalised to $\msusy$, are shown for different values of \msusy, namely: $\msusy=1\text{ TeV}$ (red), $\msusy=10\text{ TeV}$ (blue) and $\msusy=100\text{ TeV}$ (green).}
\label{fig:stop_masses}
\end{figure}

In general three experimental inputs are needed in order to constrain the three parameters $X_t$, $\mtL$ and $\mtR$, which affect the stop masses and the stop mixing angle as detailed in \cref{sec:Xt_stop_sector}. Thus, a determination of $X_t$ just from the measured values of the two stop masses is only possible if a certain relation between $\mtL$ and $\mtR$ is assumed. The dependence of the stop masses on $X_t$ is illustrated in \cref{fig:stop_masses} for the two cases  $m_{\tilde{t}_L} = m_{\tilde{t}_R} = \msusy$ (left) and $m_{\tilde{t}_L} = 0.9\, m_{\tilde{t}_R}$ (right). In both plots the lighter stop mass (solid curves) and the heavier stop mass (dashed curves), normalised by their geometric mean \msusy, are shown for $\msusy = 1 \tev$ (red), $\msusy = 10\tev$ (blue), and $\msusy = 100 \tev$ (green), and $\tan\beta = 20$ has been chosen.

In the left plot the dependence of the stop masses on $\widehat X_t$ is quite pronounced for $\msusy = 1\tev$, while the slope of the curves gets significantly smaller with larger \msusy. Especially for $\msusy = 100 \tev$ a very high mass resolution would be required to extract $X_t$ from the measured stop mass values. As is illustrated in the right plot of \cref{fig:stop_masses}, changing the assumption on the soft SUSY-breaking parameters from $\mtL = \mtR$ to $\mtL = 0.9 \, \mtR$ has a large impact. In this case, the mass difference between the two stops is rather large, while the dependence on $X_t$ is diminished. The prospects for determining $X_t$ from the stop masses would be significantly worse than for the case with $\mtL = \mtR$, in particular for \msusy\ values in excess of $10\tev$. 

These simple examples underline the obvious fact that in general it is not possible to disentangle to what extent the mass difference between the two stop masses is caused by stop mixing or by a splitting between the stop soft SUSY-breaking parameters. Accordingly, measurements of just the two stop masses alone will of course not be sufficient to determine $X_t$.


\subsection{Stop mixing angle}

As discussed in \cref{sec:Xt_stop_sector}, the stop mixing angle determines the mixing between the 
stop gauge eigenstates $\tilde t_L$ and $\tilde t_R$. 
Consequently, processes involving stops that are induced by the $SU(2)_L$ interaction are sensitive to the stop mixing angle.

At hadron colliders like the LHC, stops are predominantly produced via QCD interactions. Therefore, the branching ratios of the stops need to be disentangled and measured precisely in order to extract the stop mixing angle, which is experimentally challenging (see~\ccite{Rolbiecki:2009hk} for an exploratory study).\footnote{Another possibility would be to measure the kinematic shapes of decay processes like 
$\tilde t_{1, 2} \rightarrow t \tilde\chi$, where $\tilde\chi$ is a neutralino. This is, however, experimentally even more challenging.}

At lepton colliders, the stop mixing angle could be extracted more easily, since the stops can potentially be produced with a sizeable rate via processes involving the electroweak gauge bosons~\cite{Bartl:1997yi,Berggren:1999ss,Bartl:2000kw,Finch:2002qy}. However, stops with masses of a few TeV may be beyond the kinematic reach of the next generation of $e^+e^-$ colliders.

\begin{figure}
\centering
\includegraphics[width=.49\textwidth]{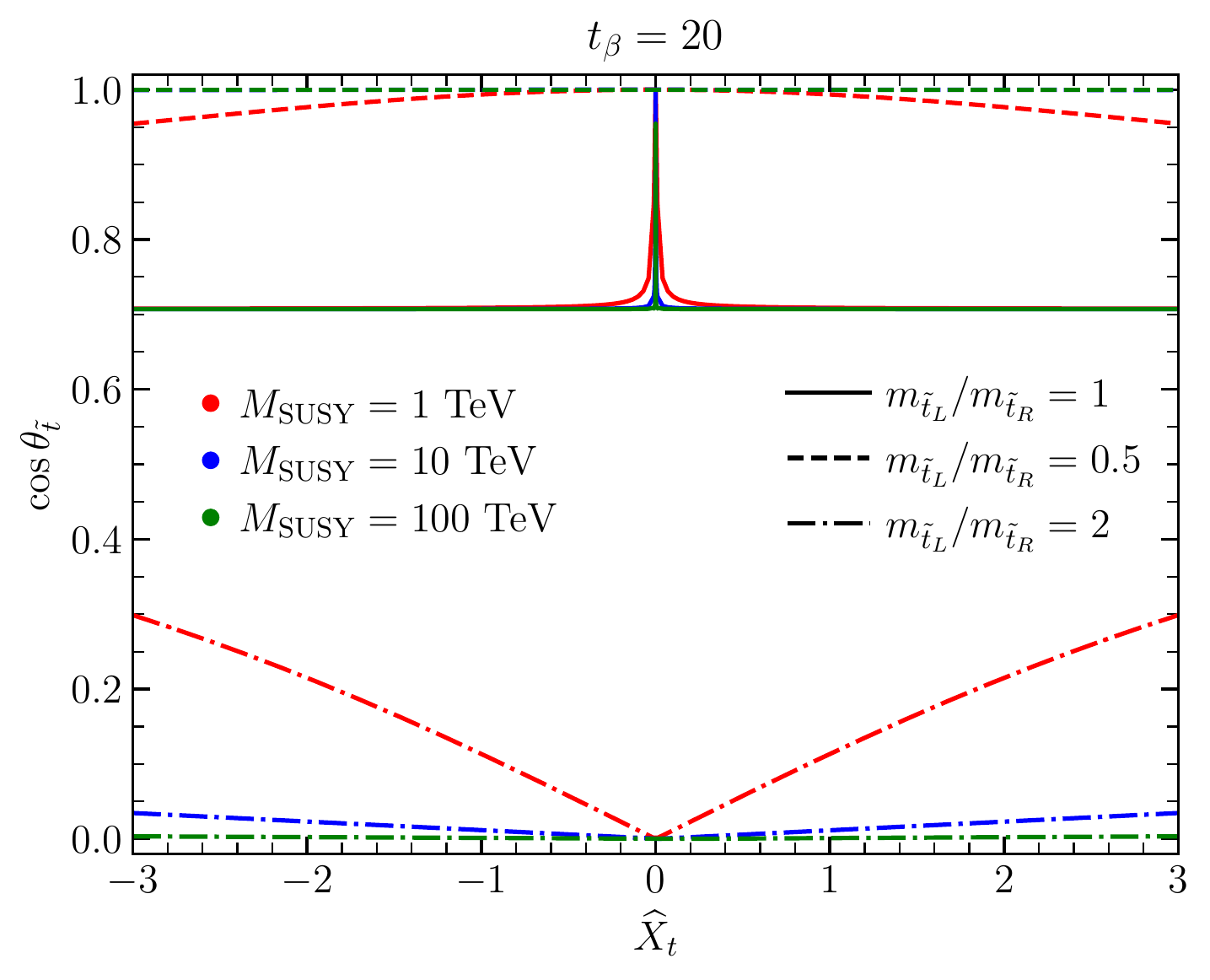}
\includegraphics[width=.495\textwidth]{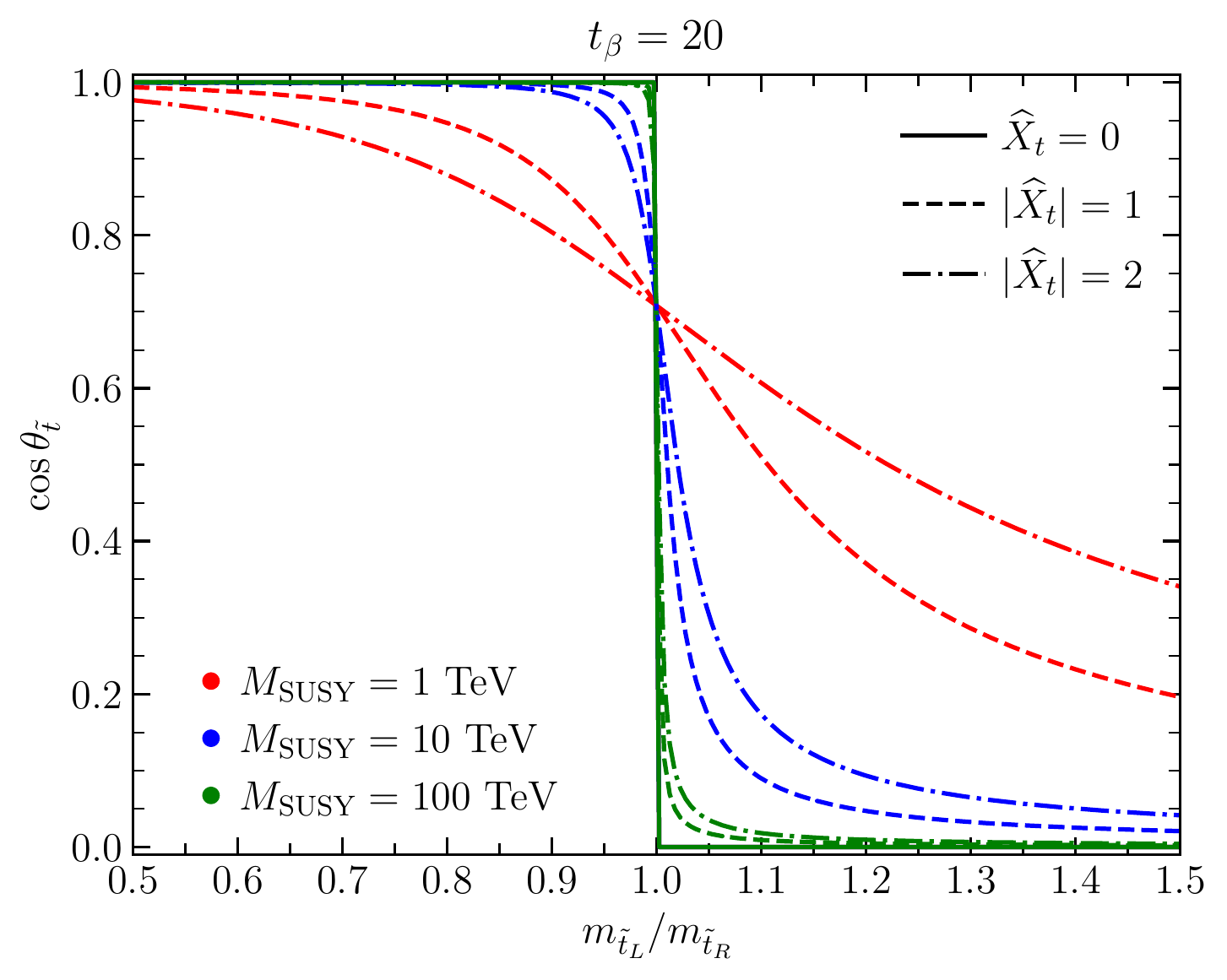}
\caption{Cosine of the stop mixing angle as a function of 
$\widehat X_t$ (\textit{left}) 
and as a function of
$m_{\tilde t_L}/m_{\tilde t_R}$ 
(\textit{right}) for different values of \msusy\ and $\tan\beta = 20$.}
\label{fig:stop_mixingangle}
\end{figure}

Even if a measurement of the stop mixing angle becomes possible, it is important to take into account that the fundamental parameter of the underlying theory is not the stop mixing angle but in fact the stop mixing parameter $X_t$. From the measured values of the two stop masses and the stop mixing angle it is in principle possible to determine the model parameters $X_t$, $m_{\tilde t_L}$ and $m_{\tilde t_R}$. However, as is illustrated in \cref{fig:stop_mixingangle}, even a precise measurement of the stop mixing angle (together with measurements of the stop masses) does not necessarily allow a reliable determination of the stop mixing parameter.

The left panel of \cref{fig:stop_mixingangle} shows the cosine of the stop mixing angle derived at the tree level as a function of $\widehat X_t$ (all interactions induced by $SU(2)_L$ interactions are proportional to $\cos\theta_{\tilde t}$). Results are shown for three different values of \msusy\ --- $1\tev$ (red), $10\tev$ (blue), $100\tev$ (green) --- and three different values of $m_{\tilde t_L}/m_{\tilde t_R}$ --- 1 (solid lines), 0.5 (dashed lines), 2 (dot-dashed lines). One can see that the determination of $X_t$ from a given input for $\cos\theta_{\tilde t}$ (and for $m_{\tilde t_1}$, $m_{\tilde t_2}$) will only be possible with a good sensitivity if the overall stop mass scale is not too large and if there is a significant splitting between the stop soft SUSY-breaking parameters. For $m_{\tilde t_L} = m_{\tilde t_R}$, the two diagonal entries of the stop mass matrix are approximately equal to each other (up to electroweak terms). In this situation, the size of the off-diagonal entries, controlled by $X_t$, has only a minor impact on the mixing angle, whose cosine is $\sim 1/\sqrt{2}$. An exception are only $X_t$ values very close to zero, where the cosine of the stop mixing angle as a function of $X_t$ develops a sharp peak towards $\cos\theta_{\tilde t} = 1$, which is reached for $X_t = 0$. This behaviour is a consequence of the electroweak contributions and the condition that $m_{\tilde t_1} \le m_{\tilde t_2}$. For increasing \msusy, the numerical impact of the electroweak terms becomes smaller and smaller, resulting in a sharper peak. In the case of a sizeable splitting between $\mtL$ and $\mtR$, displayed in our example for $\mtL/\mtR = 0.5$ and $2$, the stop mixing angle depends more sensitively on $X_t$ if \msusy is around the TeV scale. For larger \msusy the dependence becomes more and more flat. This is a consequence of the fact that the diagonal entries of the stop mass matrix are of $\mathcal{O}(\msusy^2)$ while the off-diagonal entries are of $\mathcal{O}(m_t \msusy)$.

The dependence on $m_{\tilde t_L}/m_{\tilde t_R}$ is further explored in the right panel of \cref{fig:stop_mixingangle} showing the cosine of the stop mixing angle, once again computed at tree level, as a function of $m_{\tilde t_L}/m_{\tilde t_R}$. The results are shown for three different values of \msusy\ --- $1\tev$ (red), $10\tev$ (blue), $100\tev$ (green) --- and three different values of $|\xt|$ --- 0 (solid lines), 1 (dashed lines), 2 (dot-dashed lines),\footnote{The cosine of the mixing angle does not depend on the sign of \xt\ as expected by the fact that the sign can be rewritten as a phase (see for instance \ccite{Frank:2006yh}). This is also visible in the left panel of \cref{fig:stop_mixingangle}.} while $\tan\beta$ is set to 20.

For $\xt = 0$, the stops do not mix, and the stop mixing angle does not depend on \msusy. Consequently, all solid lines lie on top of each other. The step at $m_{\tilde t_L}/m_{\tilde t_R} = 1$ is a consequence of demanding $m_{\tilde t_1} \le m_{\tilde t_2}$. While for $\msusy = 1\tev$ the different $|\xt|$ curves are still well separated for $m_{\tilde t_L}/m_{\tilde t_R} \lesssim 0.8$ and $m_{\tilde t_L}/m_{\tilde t_R} \gtrsim 1.1$, the curves approach each other for rising \msusy implying the need for more and more precise measurements of $m_{\tilde t_L}$, $m_{\tilde t_R}$ and $\cos\theta_{\tilde t}$ in order to extract $X_t$.

In summary, a precise extraction of the stop mixing parameter from the measurement of the stop mixing angle (in combination with the measurements of the two stop masses) is only possible if there is a large splitting between the soft SUSY-breaking parameters in the stop sector and if the overall stop mass scale (i.e.~the mean of the stop masses) is close to the TeV scale. If, on the other hand, the experimental information on the two stop masses and the stop mixing angle reveals that an approximate equality $m_{\tilde t_L} \approx m_{\tilde t_R}$ holds, the dependence of the stop masses on $X_t$ as discussed in \cref{fig:stop_masses} can be utilised to constrain $X_t$. However, also in this case the sensitivity to $X_t$ is significantly diminished for stop masses in the multi-TeV regime (see the discussion above). While at first sight it might seem that at a collider of sufficient energy that has the capability for precise measurements of the stop masses and the stop mixing angle
it should be possible to obtain 
a precise determination of the stop mixing parameter also for stop masses beyond the TeV scale, this is in fact not the case as a consequence of contributions that scale like $m_t/\msusy$.


\subsection{Higgs--stop--stop interaction}

Another possibility for experimentally probing $X_t$ is the investigation of observables involving the Higgs--stop--stop interaction at the tree level. In this context, the decay $\tilde t_{2} \rightarrow \tilde t_{1} +h$ appears to be most promising.\footnote{Other processes like Higgs-induced di-stop production are experimentally much harder to access.}

The associated $\tilde t_2$ decay width is given by
\begin{align}
d\Gamma_{\tilde t_2 \to \tilde t_1 h} &= \frac{1}{64\pi^2}\frac{\sqrt{(m_{\tilde t_2}^2 - (m_{\tilde t_1} + m_h)^2)(m_{\tilde t_2}^2 - (m_{\tilde t_1} - m_h)^2)}}{m_{\tilde t_2}^3}|\mathcal{M}(\tilde t_2 \rightarrow \tilde t_1 h)|^2 d \cos\theta,
\end{align}
where $\theta$ is the angle between the 3-momenta of the final state particles. The matrix element $\mathcal{M}(\tilde t_2 \rightarrow \tilde t_1 h)$ is proportional to $X_t$ (see \cref{eq:hSt1St2_coupling}).

It is instructive to discuss the limiting kinematic cases for this decay. First, we consider the case in which the heavier stop is much heavier than the SM-like Higgs boson and the lighter stop ($m_{\tilde t_1}, m_h \ll m_{\tilde t_2}$). In this case, the total decay width behaves as
\begin{align}\label{eq:W_st2_st1h_nondeg}
d\Gamma_{\tilde t_2\to\tilde t_1 h} \xrightarrow{m_h, m_{\tilde{t}_1} \ll m_{\tilde{t}_2}} \frac{1}{64\pi^2}\frac{1}{m_{\tilde{t}_2}}|\mathcal{M}(\tilde t_2 \rightarrow \tilde t_1 h)|^2d\cos\theta\propto \frac{|X_t|^2}{m_{\tilde{t}_2}},
\end{align}
where in the final step we only consider mass scales to obtain an estimate of the behaviour of the decay width. Assuming that $|X_t| \sim \mathcal{O}(\msusy)$, then also $\Gamma_{\tilde t_2\to\tilde t_1 h}$ is of $\mathcal{O}(\msusy)$.

As a second limiting case, we consider the situation in which the two stops are approximately mass-degenerate and much heavier than the SM-like Higgs boson ($m_h \ll m_{\tilde{t}_1} \sim m_{\tilde{t}_2}$). In this limit the total decay width becomes
\begin{align}\label{eq:W_st2_st1hdeg}
d\Gamma_{\tilde t_2\to\tilde t_1 h} \xrightarrow{m_h \ll m_{\tilde{t}_1} \sim m_{\tilde{t}_2}}& \frac{1}{64\pi^2}\frac{|m_{\tilde t_2}^2 - m_{\tilde t_1}^2|}{m_{\tilde{t}_2}^3}|\mathcal{M}(\tilde t_2 \rightarrow \tilde t_1 h)|^2 d\cos\theta\simeq \nonumber\\
\simeq{}& \frac{1}{64\pi^2}\frac{2 m_t |X_t|}{m_{\tilde{t}_2}^3}|\mathcal{M}(\tilde t_2 \rightarrow \tilde t_1 h)|^2 d\cos\theta \propto \frac{m_t |X_t|^3}{m_{\tilde{t}_2}^3}.
\end{align}
Comparing to \cref{eq:W_st2_st1h_nondeg}, the decay width is suppressed by a factor of $m_t |X_t|/m_{\tilde t_2}^2$ in comparison to the case with $m_{\tilde t_1}, m_h \ll m_{\tilde t_2}$.\footnote{As a consequence of expanding around $m_h/m_{\tilde t_{1,2}}$, we implicitly assumed the $h$ boson to be massless, so that $m_h$ does not appear in \cref{eq:W_st2_st1hdeg}. } 
Assuming again that $|X_t| \sim \mathcal{O}(\msusy)$, $\Gamma_{\tilde t_2\to\tilde t_1 h}$ is of $\mathcal{O}(m_t)$. This means that the decay width is much smaller than in the case of $m_h \sim m_{\tilde t_1} \ll m_{\tilde t_2}$, 
which is mainly due to the suppression of the phase space. Consequently, other decay channels of $\tilde t_2$ can easily dominate over the $\tilde t_2 \to \tilde t_1 + h$ decay channel in this case.

\begin{figure}
\centering
\begin{minipage}{.48\textwidth}
\includegraphics[width=\textwidth]{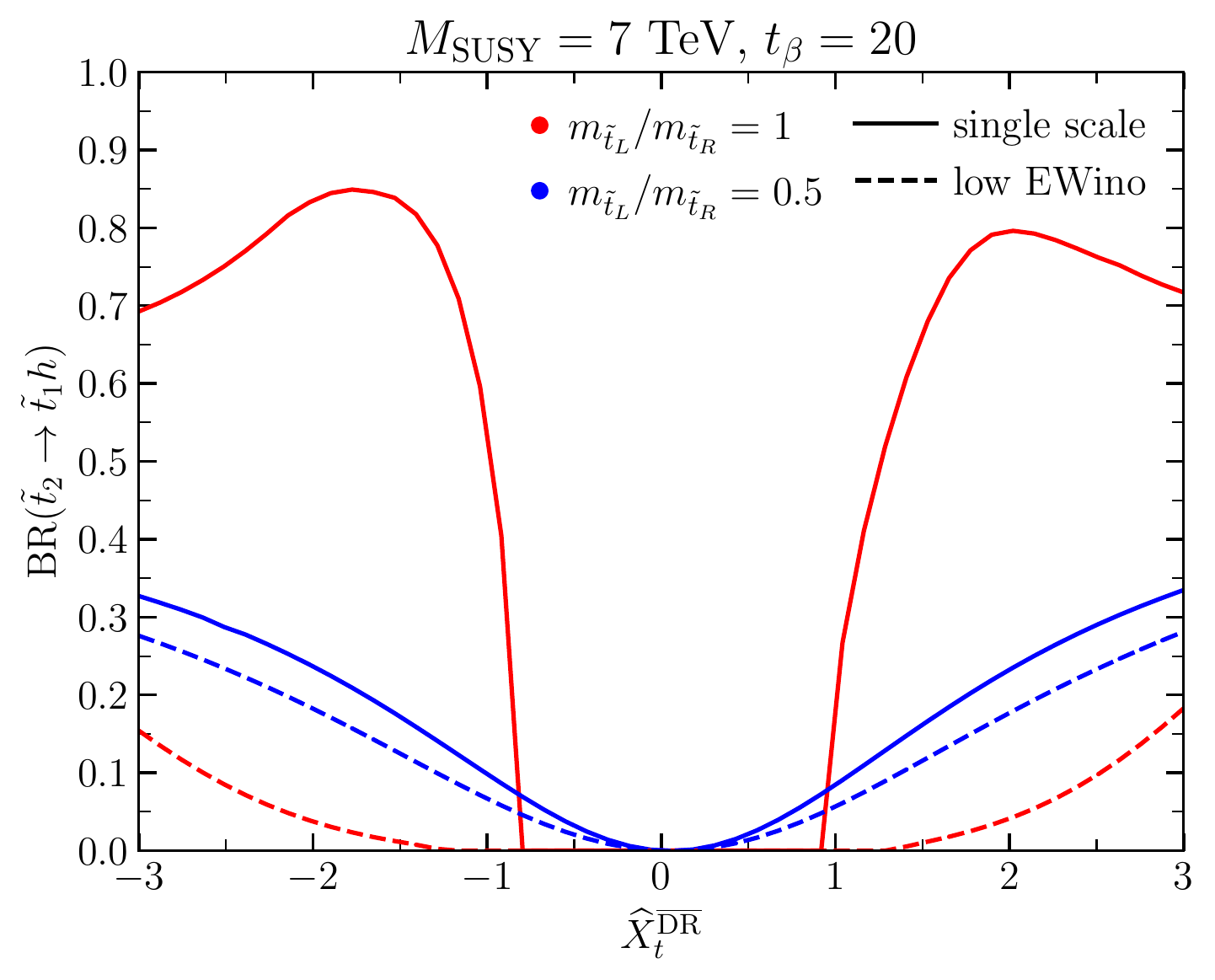}
\end{minipage}
\begin{minipage}{.48\textwidth}
\includegraphics[width=\textwidth]{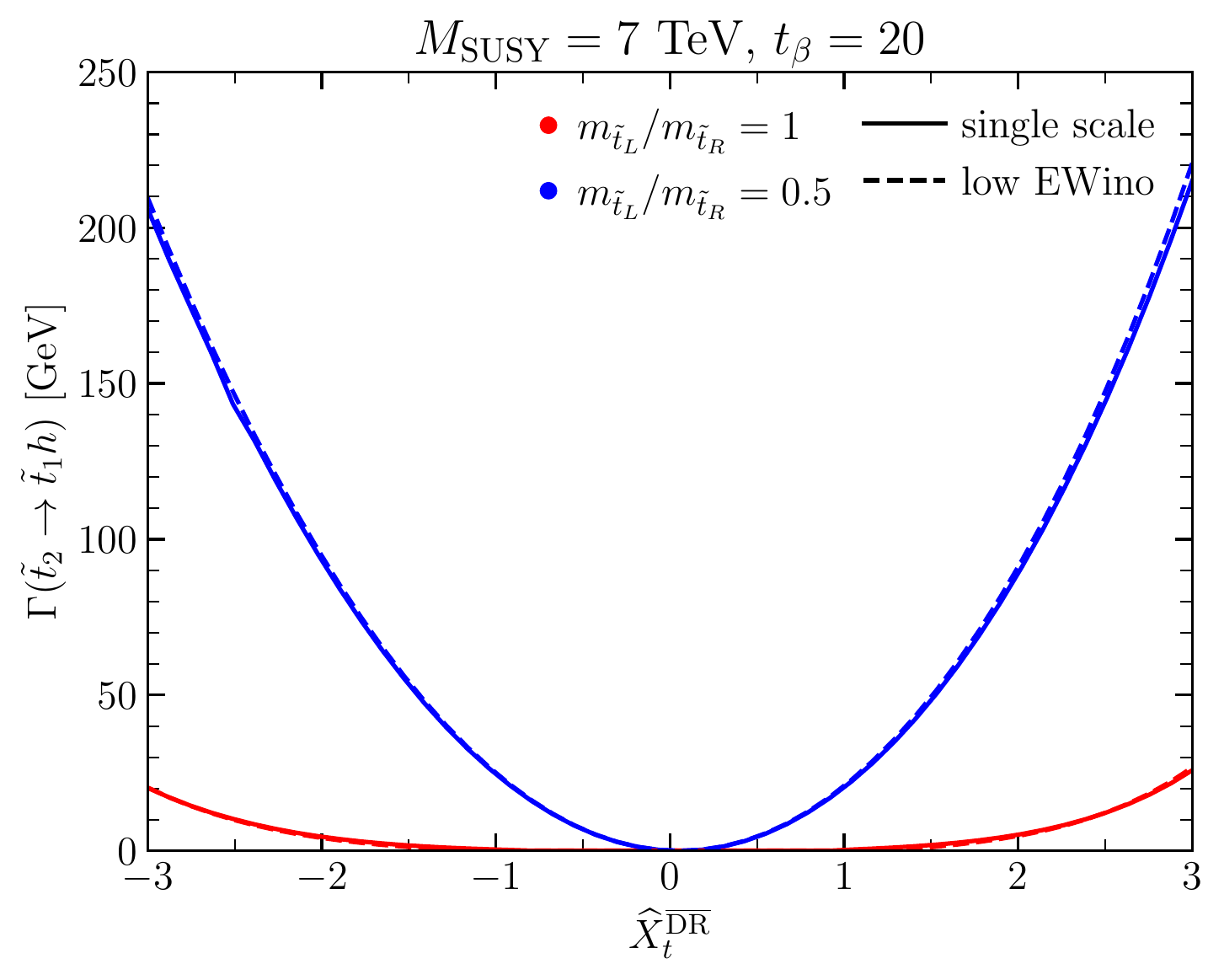}
\end{minipage}
\caption{\textit{Left}: branching ratio of $\tilde t_2$ into $\tilde t_1$ and a SM-like Higgs boson as a function of $\widehat X_t^\DR$ for $m_{\tilde t_L}/m_{\tilde t_R} = 1$ (red lines) and $m_{\tilde t_L}/m_{\tilde t_R} = 0.5$ (blue lines). Results are shown for a single-scale scenario (solid lines) and a scenario with light EWinos (dashed lines). \textit{Right}: same as left panel, but the $\tilde t_2 \to \tilde t_1 + h$ partial decay width is shown. The dashed lines lie on top of the solid lines.}
\label{fig:stop_br}
\end{figure}

We numerically investigate this behaviour in \cref{fig:stop_br} showing the branching ratio of $\tilde t_2 \to \tilde t_1 + h$ (left panel) and the $\tilde t_2\to \tilde t_1 + h$ partial decay width (right panel) as a function of $\xt$, which for this Figure is chosen to be renormalised in the \DR scheme, see below. The branching ratio and decay width are evaluated using \texttt{SUSY-HIT}~\cite{Djouadi:1997yw,Djouadi:2002ze,Muhlleitner:2003vg,Djouadi:2006bz}, which includes the leading QCD corrections to $\tilde t_2 \to \tilde t_1 + h$, for two different scenarios: in the single scale scenario (solid curves), all soft SUSY-breaking masses 
(except for the stop SUSY-breaking masses) as well as the $\mu$ parameter and the mass scale of the heavy Higgs bosons are chosen to be equal to $\msusy = 7 \tev$ (and $t_\beta = 20$ is used); for the low mass electroweakino scenario (dashed curves), the same parameters as in the single-scale scenario are used apart from the Wino and Bino soft SUSY-breaking masses ($M_1$ and $M_2$) as well as $\mu$ which are chosen to be equal to $\msusy/2$ (implying the existence of comparably lighter neutralinos and charginos). For the red curves, the stop soft-breaking masses are chosen to be equal to \msusy, while for the blue curves they are set to $\mtL = \msusy/\sqrt{2}$ and $\mtR = \sqrt{2}\msusy$.

We observe the largest branching ratio of $\sim 70-85\%$ in the single scale scenario with mass-degenerate stop soft-breaking masses for $|\xt| > 1.5$ (see the left panel of \cref{fig:stop_br}). In this case, the $\tilde t_2 \to \tilde t_1 + h$ decay channel is enhanced by a factor $|\xt|^3$ in comparison to the $\tilde t_2 \to \tilde t_1 + Z$ decay channel, the only other decay channel with a sizeable branching ratio. The $\tilde t_2 \to \tilde t_1 + h$ branching ratio, however, quickly goes to zero if $|\xt|$ approaches zero, since in this limit the stop masses become equal and the phase space of the $\tilde t_2 \to \tilde t_1 + h$ decay vanishes.

As expected from \cref{eq:W_st2_st1hdeg}, the $\tilde t_2 \to \tilde t_1 + h$ decay width is quite small ($\lesssim 20\gev$) in the single scale scenario with mass-degenerate stop soft-breaking masses (see the right panel of \cref{fig:stop_br}). If low-mass electroweakinos are present, the $\tilde t_2 \to \tilde t_1 + h$ decay width is unchanged. As a consequence of the small $\tilde t_2 \to \tilde t_1 + h$ decay width, the presence of additional decay channels for $\tilde t_2$ (i.e., $\tilde t_2\to t\tilde\chi^0, b\tilde\chi^+$), however, suppresses the branching ratio of the $\tilde t_2\to \tilde t_1 + h$ decay channel to values below $\sim 15\%$.

The situation is different if $m_{\tilde t_1} \ll m_{\tilde t_2}$ (see blue curves in \cref{fig:stop_br}). In this case, the $\tilde t_2 \to \tilde t_1 + h$ decay width is significantly larger ($\lesssim 200\gev$) as expected from \cref{eq:W_st2_st1h_nondeg}. This mass hierarchy, however, also allows large partial decay widths for other decay channels like $\tilde t_2\to \tilde t_1 + Z$. Consequently, $\text{BR}(\tilde t_2 \to \tilde t_1 + h)$ reaches only maximal values of $\sim 33\%$ for $\xt \sim \pm 3$ in this scenario. In contrast to the case of $\mtL=\mtR$, the presence of low-mass electroweakinos further lowers this branching ratio by only $\lesssim 5\%$, since for $m_{\tilde t_1} \ll m_{\tilde t_2}$ the partial decay widths of the $\tilde t_2 \to \tilde t_1 + h$ decay and the electroweakino decays are similar in size (as expected from \cref{eq:W_st2_st1h_nondeg}). 

\medskip

In summary, the usefulness of the $\tilde t_2 \rightarrow \tilde t_1 + h$ process to extract $X_t$ crucially depends on the sparticle mass hierarchy. The presence of additional decay channels or an approximate mass degeneracy between the stop quarks can easily suppress the $\tilde t_2\to\tilde t_1 + h$ branching ratio making it hard to measure it precisely at a future experiment. 


\subsection{Relation to the mass and the couplings of the SM-like Higgs boson}

In view of the discussion above, one might wonder whether for stop masses in the multi-TeV regime the parameter $X_t$ has any significant phenomenological impact at all. However, the situation is very different regarding the impact of the parameter $X_t$ on the prediction for the mass of the SM-like Higgs boson.

While the Higgs-boson mass is a free parameter in the SM, the mass of the SM-like Higgs boson in the MSSM, $M_h$, can be computed in terms of the model parameters as a consequence of the underlying symmetry (see \ccite{Slavich:2020zjv} for a recent review). $M_h$ is bounded to be below $M_Z$ at the tree level. Loop corrections can, however, increase it to the experimentally measured value of $\simeq 125\gev$.

The dominant corrections at the one-loop level arise from the stop/top sector and are, in the limit $M_A\gg M_Z$ ($M_A$ being the $A$ boson mass), controlled by the two parameters \msusy and $X_t$,
\begin{align}
M_h^2 \simeq m_h^2 + 12 k \frac{m_t^4}{v^2} \left(\ln\frac{\msusy^2}{m_t^2} + |\widehat X_t|^2 - \frac{1}{12}|\widehat X_t|^4\right) + \ldots ,
\end{align}
where $m_h^2 = M_Z^2 c_{2\beta}^2$ is the tree-level mass (again in the limit $M_A\gg M_Z$), \msusy is the geometric mean of the stop masses, $k\equiv 1/(16\pi^2)$ is the loop factor, and the ellipsis denotes subdominant one-loop and higher-order terms. Apart from \msusy and $t_\beta$ (and SM parameters),\footnote{The mass scale of the heavy Higgs bosons enters at lowest order and can in principle have a significant impact on the prediction for $M_h$. Existing search limits in combination with measurements of the properties of the SM-like Higgs boson, however, put strong lower bounds on $M_A$. Since for $M_A \gg M_Z$ the increase in $M_h$ for rising $M_A$ quickly saturates, in the phenomenologically viable mass region of $M_A$ (see e.g.\ \ccite{Bagnaschi:2018ofa,Bahl:2019ago,ATLAS:2020naq}) the dependence of $M_h$ on $M_A$ is subdominant compared to the dependence on $\msusy$ and $X_t$.} only \xt strongly influences the dominant one-loop correction. This means that if both stops are discovered, the measurement of $M_h$ can be used to determine $X_t$ in the MSSM (under the assumption of a specific $t_\beta$ value).
    
\begin{figure}
\centering
\includegraphics[width=.6\textwidth]{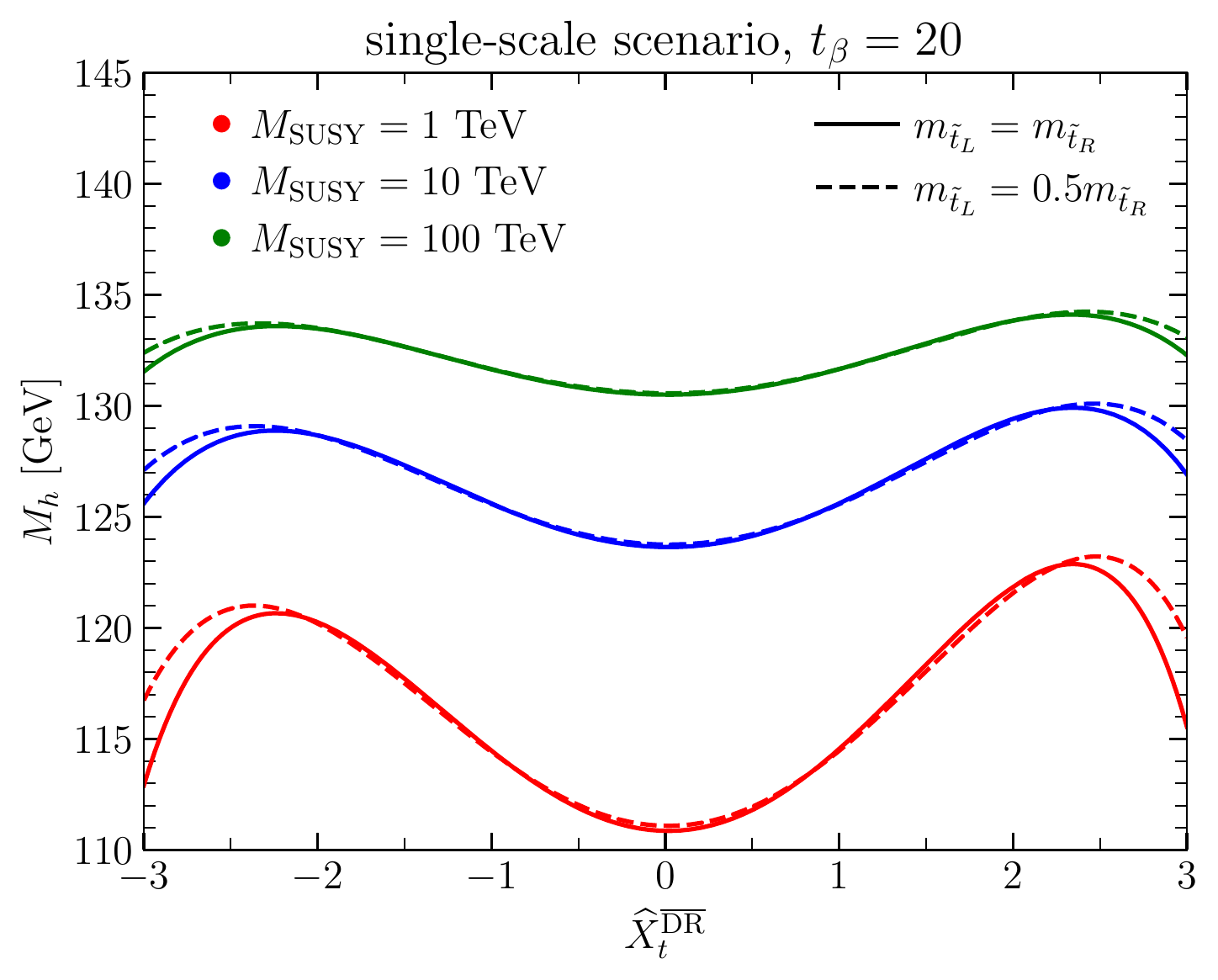}
\caption{The mass of the SM-like Higgs boson, $M_h$, as a function of $\widehat X_t$. Results for $M_h$ are shown in a single-scale scenario with $t_\beta = 20$, for two possible choices of the stop soft-SUSY breaking masses, namely $m_{\tilde{t}_L}=m_{\tilde{t}_R}$ (solid curves) and $m_{\tilde{t}_L}=0.5\ m_{\tilde{t}_R}$ (dashed curves), and for three different values of $\msusy$: $\msusy=1\text{ TeV}$ (red), $\msusy=10\text{ TeV}$ (blue), and $\msusy=100\text{ TeV}$ (green).}
\label{fig:Mh_a}
\end{figure}
    
\begin{figure}
\centering
\begin{minipage}{.48\textwidth}
\includegraphics[width=\textwidth]{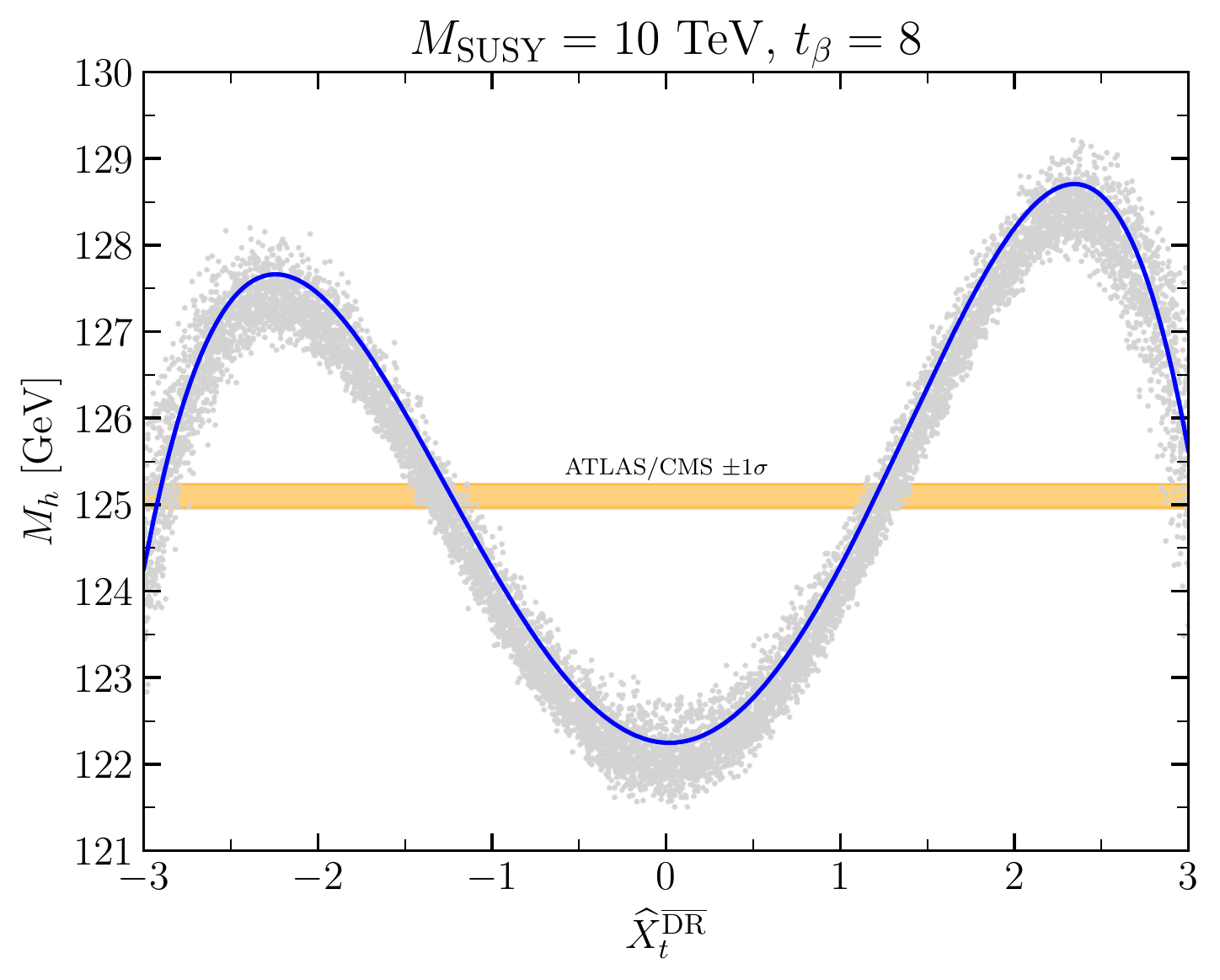}
\end{minipage}
\begin{minipage}{.48\textwidth}
\includegraphics[width=\textwidth]{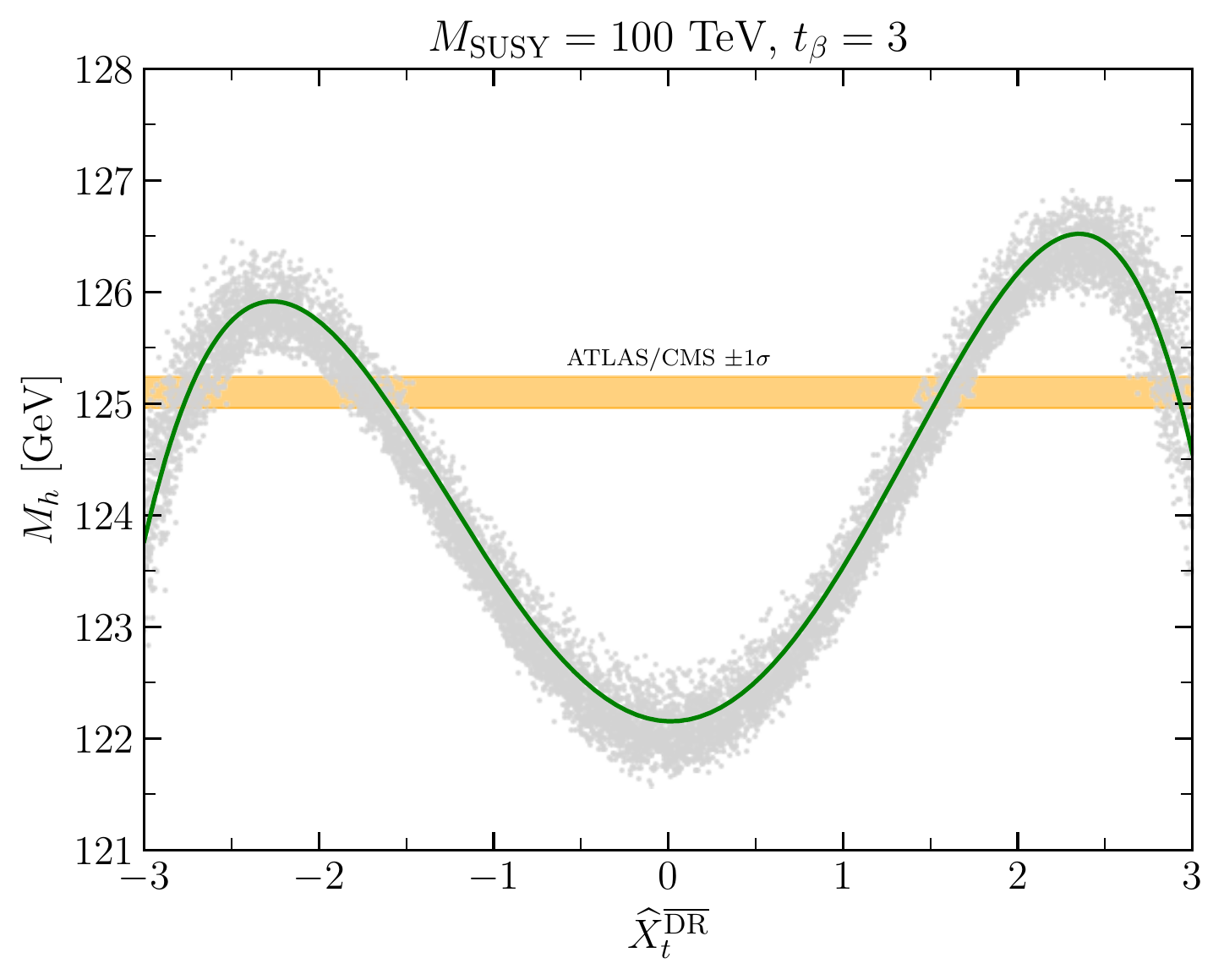}
\end{minipage}
\caption{The mass of the SM-like Higgs boson, $M_h$, as a function of $\widehat X_t$. \textit{Left}: The blue curve displays the predicted value of $M_h$ for $t_\beta=8$ in a single-scale scenario with $\msusy=10\tev$, where all BSM mass terms are set to $\msusy$ and all trilinear couplings other than $A_t$ are set to zero. The gray points are obtained by varying the mass parameters and the trilinear couplings randomly in the range $[1/2 \msusy, 2\msusy]$. The orange band shows the value of the combined ATLAS/CMS measurement for $M_h$ together with its $1\sigma$ uncertainty. \textit{Right}: As for the left plot, but the green curve displays the predicted value of $M_h$ for $t_\beta=3$ and $\msusy = 100\tev$.}
\label{fig:Mh_b}
\end{figure}

This is illustrated in \cref{fig:Mh_a} showing $M_h$ as a function of $\widehat X_t^\DR$ for $t_\beta = 20$ (calculated using \texttt{FeynHiggs 2.18.1}~\cite{Heinemeyer:1998yj,Heinemeyer:1998np,Hahn:2009zz,Degrassi:2002fi,Frank:2006yh,Hahn:2013ria,Bahl:2016brp,Bahl:2017aev,Bahl:2018qog}).\footnote{\texttt{FeynHiggs} computes $M_h$ including the full one-loop corrections as well as the dominant two-loop corrections in the limit of vanishing electroweak gauge couplings. Moreover, leading, next-to-leading, and next-to-next-to-leading logarithmic contributions are resummed using an effective field theory approach. For our numerical analysis, all trilinear couplings (except $A_t$) are chosen to be zero, and the stop sector is renormalised in the \DR scheme (see \cref{sec:stop_DR_MDR_schemes}).} As above, \msusy is defined to be the geometric mean of the two stop masses. A single-scale scenario is considered, where all soft SUSY-breaking masses (as well as the $A$ boson mass $M_A$ and $\mu$) are chosen to be equal to \msusy which is set to $1\tev$ (red), $10\tev$ (blue), and $100\tev$ (green). For the solid lines, $m_{\tilde t_L} = m_{\tilde t_R}$ is set; for the dashed lines,  $m_{\tilde t_L} = 0.5\ m_{\tilde t_R}$ is used. For $\msusy = 1\tev$, the prediction for $M_h$ shows a very pronounced dependence on $\widehat X_t^\DR$ varying between $\sim 111\gev$ and $\sim 123\gev$ within the considered range of $\widehat X_t^\DR$.\footnote{Outside the range $-3 \lesssim \widehat X_t \lesssim 3$, colour-breaking minima can occur rendering this region in large parts unphysical (see e.g.~\ccite{Bechtle:2016kui,Hollik:2018wrr}).} The sizeable variation of the prediction for $M_h$ with $\widehat X_t^\DR$ even for the much higher values for \msusy of $10\tev$ (blue curves) and $100\tev$ (green curves) indicates the potential for a precise determination of $X_t$ from the achieved high-precision measurement of $M_h$ (for scenarios where the theoretical prediction is compatible with the experimental value and assuming further progress on the reduction of the theoretical uncertainty in the prediction of $M_h$, which for instance in a scenario in which all BSM particles are close in mass has been estimated to be $\lesssim 1\gev$, see the discussion in \ccite{Bahl:2019hmm}). Even if the assumption $\mtL = \mtR$ is relaxed and e.g.\ $\mtL = 0.5\ \mtR$ (dashed lines) is chosen, the prediction for $M_h$ only changes significantly for $|\widehat X_t| \gtrsim 2.5$, showing the robustness of the dependence of $M_h$ on $X_t$.

These findings are further demonstrated in \cref{fig:Mh_b}. In the left panel, we show the dependence of $M_h$ on $\widehat X_t^\DR$ for $\msusy = 10\tev$ and $t_\beta = 8$\footnote{Qualitatively similar results are expected for $\tan\beta=20$, as in \cref{fig:Mh_a}.} (blue curve) in comparison to the current $1\,\sigma$ experimental uncertainty band in orange. The theoretical prediction for $M_h$ as a function of $\widehat X_t^\DR$ has the following sources of uncertainties: unknown higher-order corrections, see \ccite{Bahl:2019hmm} for a detailed discussion of the theoretical uncertainty of the Higgs mass calculation implemented in \texttt{FeynHiggs}, the experimental errors of the input parameters of the SM, see \ccite{Slavich:2020zjv}, and the lacking knowledge of the values of the other SUSY parameters entering the prediction for $M_h$. In order to illustrate the effect of the latter uncertainty, in addition to the parameter setting of the blue curve, for which all non-SM masses are set equal to \msusy and all trilinear couplings (apart from $A_t$) are set to zero, we randomly vary each of these parameters (including all trilinear couplings except $A_t$, which is fixed via $X_t$) independently in the interval $[1/2 \msusy, 2 \msusy]$. In this way, we have produced $10^4$ parameter points. They are shown in the form of small grey points in addition to the blue curve. We find that all of these points lie within $\sim 0.5\gev$ of the single scale scenario. This implies that in a situation where the stop masses and $t_\beta$ will be known with reasonable accuracy in the future a reliable determination of $X_t$ would be possible from confronting the theoretical prediction for $M_h$ with the experimental value even if the information about the other parts of the SUSY spectrum is very limited (see also discussion in \ccite{Djouadi:2013lra}). A similar approach that semi-analytically expresses the stop trilinear coupling in terms of $M_h$ has recently been discussed in \ccite{ElKosseifi:2022rkf}.

The right panel of \cref{fig:Mh_b} displays the remarkable feature that an indirect determination of $X_t$ from $M_h$ can even be achieved for a SUSY scale as high as $\msusy = 100 \tev$. The green curve shows the prediction for $M_h$ for $\msusy = 100\tev$ and $t_\beta = 3$. As in the left plot, the small grey points show the random variations of the SUSY parameters in the interval $[1/2 \msusy, 2 \msusy]$. The fact that the prediction for $M_h$ has a high sensitivity to the ratio $X_t/\msusy$ even for the case where the SUSY scale is so high that it may be beyond the reach of any future collider clearly shows the unique role of the prediction for the SM-like Higgs mass in constraining the stop mixing parameter $X_t$ (of course, from the information about $M_h$ alone $X_t$ cannot be fully determined).

While only the prediction for the mass of the SM-like Higgs boson shows the special feature that it has a high sensitivity to $X_t/\msusy$ even for very high values of $\msusy$, the effects of varying $X_t$ on the couplings of the SM-like Higgs boson tend to vanish in the decoupling region where \msusy is large. On the other hand, for relatively low SUSY scales also high-precision measurements of the branching ratios of the SM-like Higgs boson can provide supplementary information on $X_t$. Exploratory studies addressing the sensitivity of Higgs branching ratios to $X_t$ have been carried out in \ccite{Desch:2004cu,LHCLCStudyGroup:2004iyd}.

%% file: sec_Xt_renormalization.tex
Based on the discussion of the various approaches to determine $X_t$ from physical observables, we compare in this Section different renormalisation schemes for the top/stop sector. 

Working for simplicity in the limit of vanishing electroweak gauge coupling, we renormalise the parameters appearing in the stop mass matrix (see \cref{eq:stop_mass_matrix}) as follows,\footnote{See \ccite{Fritzsche:2011nr, Hollik:2014wea, Hollik:2014bua, Passehr:2014xwu} for a detailed discussion, employing similar notations, of the renormalisation scheme presented in this Section. Further discussions of the renormalisation of the stop/top sector can be found in \ccite{Eberl:1996np,Djouadi:1996wt,Beenakker:1996de,Djouadi:1998sq,Bartl:1997pb,Bartl:1998xp,Guasch:1998as,Kraml:1999qd,Brignole:2001jy,Hollik:2003jj,Heinemeyer:2007aq,Baro:2009gn,Fritzsche:2013fta}.}
\begin{eqnarray}
  \label{eq:mstLR_Xt_mt_CT_transformation}  
  \begin{aligned}
    & m_{\tilde{t}_{L/R}}^2 \to m_{\tilde{t}_{L/R}}^2 + \deltaOL m_{\tilde{t}_{L/R}}^2, \\
    & X_t \to X_t + \deltaOL X_t, \\
    & m_t \to m_t + \deltaOL m_t.
  \end{aligned}
\end{eqnarray}
In this way the stop mass matrix $\mathbf{M}_{\tilde{t}}$ acquires the counterterm
\begin{equation}
  \label{eq:stop_mass_matrix_CT}
  \deltaOL \mathbf{M}_{\tilde{t}} =
  \begin{pmatrix}
    \deltaOL m_{\tilde{t}_L}^2 + \deltaOL m_{t}^2 & X_{t}^* \; \deltaOL m_{t} + m_{t} \; \deltaOL X_{t}^*  \\
    X_{t} \; \deltaOL m_{t} + m_{t} \; \deltaOL X_{t} & \deltaOL m_{\tilde{t}_R}^2 + \deltaOL m_{t}^2
  \end{pmatrix} \; . 
\end{equation}
Using the tree-level transformation matrix $\mathbf{U}_{\tilde{t}}$, which relates gauge and mass eigenstates (see~\cref{eq:stop_LR_12_transformation}), we define
\begin{equation}
  \label{ch3:eq67}
  \mathbf{U}_{\tilde{t}} \; \deltaOL \mathbf{M}_{\tilde{t}} \; \mathbf{U}_{\tilde{t}}^{\dagger} =
  \begin{pmatrix}
    \deltaOL m_{\tilde{t}_{1}}^2 & \deltaOL m_{\tilde{t}_{12}}^2 \\ 
    \deltaOL m_{\tilde{t}_{21}}^2 & \deltaOL m_{\tilde{t}_{2}}^2  
  \end{pmatrix} \; , 
\end{equation}
where $\deltaOL m_{\tilde{t}_{21}}^2 = (\deltaOL m_{\tilde{t}_{12}}^2)^*$.

Rotating back to the gauge-eigenstate basis, the counterterms for the soft-breaking parameters read,
\begin{subequations}
\label{eq:stop_CT_relations}
  \begin{align}
    \label{eq:stop_CT_relations_a}
    & \deltaOL X_t = \frac{1}{m_t} \left[\mathbf{U}_{\tilde{t}_{11}}\mathbf{U}_{\tilde{t}_{12}}^* \left(\deltaOL m_{\tilde{t}_{1}}^2 - \deltaOL m_{\tilde{t}_{2}}^2 \right) \right. \notag \\
      & \hspace{1.5cm} \left. + \; \deltaOL m_{\tilde{t}_{12}}^2 \mathbf{U}_{\tilde{t}_{21}}\mathbf{U}_{\tilde{t}_{12}}^* + \deltaOL m_{\tilde{t}_{21}}^2 \mathbf{U}_{\tilde{t}_{11}}\mathbf{U}_{\tilde{t}_{22}}^* - X_t \deltaOL m_t \right], \\
    \label{eq:stop_CT_relations_b}
    & \deltaOL m_{\tilde{t}_L}^2 = \deltaOL m_{\tilde{t}_{1}}^2 \vert \mathbf{U}_{\tilde{t}_{11}} \vert^2 + \deltaOL m_{\tilde{t}_{2}}^2 \vert \mathbf{U}_{\tilde{t}_{12}} \vert^2 \notag \\
    & \hspace{1.5cm} + \; \deltaOL m_{\tilde{t}_{12}}^2 \mathbf{U}_{\tilde{t}_{21}} \mathbf{U}_{\tilde{t}_{11}}^* + 
    \deltaOL m_{\tilde{t}_{21}}^2 \mathbf{U}_{\tilde{t}_{11}} \mathbf{U}_{\tilde{t}_{21}}^*
    - 2 m_t \; \deltaOL m_t, \\
    \label{eq:stop_CT_relations_c}
    & \deltaOL m_{\tilde{t}_R}^2 = \deltaOL m_{\tilde{t}_{1}}^2 \vert \mathbf{U}_{\tilde{t}_{12}} \vert^2 + \deltaOL m_{\tilde{t}_{2}}^2 \vert \mathbf{U}_{\tilde{t}_{22}} \vert^2 \notag \\
    & \hspace{1.5cm} + \; \deltaOL m_{\tilde{t}_{12}}^2 \mathbf{U}_{\tilde{t}_{22}} \mathbf{U}_{\tilde{t}_{12}}^* + 
    \deltaOL m_{\tilde{t}_{21}}^2 \mathbf{U}_{\tilde{t}_{12}} \mathbf{U}_{\tilde{t}_{22}}^*
    - 2 m_t \; \deltaOL m_t \; .
  \end{align}
\end{subequations}
At this point, a remark should be made about the renormalisation of the off-diagonal entries of the matrix $\mathbf{M}_{\tilde{t}}$. We have used $X_t$ as a free parameter, while the entries of the transformation matrix $\mathbf{U}_{\tilde{t}}$ were set to their tree-level values. Sometimes, a slightly different approach is used \cite{Heinemeyer:2007aq}. Namely, instead of renormalising the $X_t$ parameter, the angle $\theta_{\tilde{t}}$ and the phase $\phi_{X_t}$ of the rotation matrix $\mathbf{U}_{\tilde{t}}$ are renormalised,
\begin{equation}
  \label{ch3:eq71}   
  \theta_{\tilde{t}} \to \theta_{\tilde{t}} + \deltaOL \theta_{\tilde{t}}, \quad  \phi_{X_t} \to \phi_{X_t} + \deltaOL \phi_{X_t}.
\end{equation}
At the first step, the original mass matrix is expressed in terms of $\theta_t$ and $\phi_{X_t}$ as,
\begin{equation} 
  \mathbf{M}_{\tilde{t}} =
  \begin{pmatrix}
    \cos^2 \theta_{\tilde{t}} \; m_{\tilde{t}_{1}}^2 +  \sin^2 \theta_{\tilde{t}} \; m_{\tilde{t}_{2}}^2 &
    (m_{\tilde{t}_{1}}^2 - m_{\tilde{t}_{2}}^2) \sin \theta_{\tilde{t}} \; \cos \theta_{\tilde{t}} \; e^{-i\phi_{X_t}} \\
    (m_{\tilde{t}_{1}}^2 - m_{\tilde{t}_{2}}^2) \sin \theta_{\tilde{t}} \; \cos \theta_{\tilde{t}} \; e^{i\phi_{X_t}} &
    \cos^2 \theta_{\tilde{t}} \; m_{\tilde{t}_{2}}^2 +  \sin^2 \theta_{\tilde{t}} \; m_{\tilde{t}_{1}}^2
  \end{pmatrix} \; .
\end{equation}
Using the definition of the counterterms, given in \cref{eq:stop_CT_relations}, the counterterms for the entries of the original mass matrix can then be written as,
\begin{subequations}
  \begin{align}
    & \deltaOL \mathbf{M}_{\tilde{t}_{11}} = \cos^2 \theta_{\tilde{t}} \; \deltaOL m_{\tilde{t}_{1}}^2 +  \sin^2 \theta_{\tilde{t}} \; \deltaOL m_{\tilde{t}_{2}}^2
    + (m_{\tilde{t}_{2}}^2 - m_{\tilde{t}_{1}}^2) \sin 2 \theta_{\tilde{t}} \; \deltaOL \theta_{\tilde{t}}, \\ 
    \label{ch3:eq73b}
    & \deltaOL \mathbf{M}_{\tilde{t}_{12}} = (\deltaOL m_{\tilde{t}_{1}}^2 - \deltaOL m_{\tilde{t}_{2}}^2) \sin \theta_{\tilde{t}} \cos \theta_{\tilde{t}} \; e^{-i\phi_{X_t}} \notag \\
    & \hspace{1.85cm} + (m_{\tilde{t}_{1}}^2 - m_{\tilde{t}_8{2}}^2) (\deltaOL \theta_{\tilde{t}} \; \cos 2 \theta_{\tilde{t}}
    - i \deltaOL \phi_{X_t} \sin \theta_{\tilde{t}} \; \cos \theta_{\tilde{t}}) \; e^{-i\phi_{X_t}}, \\
    \label{ch3:eq73c}
    & \deltaOL \mathbf{M}_{\tilde{t}_{21}} = (\deltaOL m_{\tilde{t}_{1}}^2 - \deltaOL m_{\tilde{t}_{2}}^2) \sin \theta_{\tilde{t}} \cos \theta_{\tilde{t}} \; e^{i\phi_{X_t}} \notag \\
    & \hspace{1.85cm} + (m_{\tilde{t}_{1}}^2 - m_{\tilde{t}_{22}}^2) (\deltaOL \theta_{\tilde{t}} \; \cos 2 \theta_{\tilde{t}}
    + i \deltaOL \phi_{X_t} \sin \theta_{\tilde{t}} \; \cos \theta_{\tilde{t}}) \; e^{i\phi_{X_t}}, \\
    \label{ch3:eq73d}    
    & \deltaOL \mathbf{M}_{\tilde{t}_{22}} = \cos^2 \theta_{\tilde{t}} \; \deltaOL m_{\tilde{t}_{2}}^2 +  \sin^2 \theta_{\tilde{t}} \; \deltaOL m_{\tilde{t}_{1}}^2 
    + (m_{\tilde{t}_{1}}^2 - m_{\tilde{t}_{2}}^2) \sin 2 \theta_{\tilde{t}} \; \deltaOL \theta_{\tilde{t}} \;.
  \end{align}
\end{subequations}
By transforming the counterterm matrix, $\deltaOL \mathbf{M}_{\tilde{t}}$, to the mass eigenstates basis, we arrive at the following expression,
\begin{equation}
  \label{eq:dmst12_dthetat_dpxt_relation}
  \deltaOL m_{\tilde{t}_{12}}^2 = e^{-i\phi_{X_t}} (m_{\tilde{t}_{1}}^2 - m_{\tilde{t}_{2}}^2) (\deltaOL \theta_{\tilde{t}} - i \deltaOL \phi_{X_t} \sin \theta_{\tilde{t}} \; \cos \theta_{\tilde{t}}),
\end{equation}
which relates the off-diagonal stop mass matrix counterterm to the stop mixing angle and stop phase counterterms.


\subsection{\texorpdfstring{\DR}{DRbar} scheme}
\label{sec:stop_DR_scheme}

From a technical point of view, renormalising the stop masses and $X_t$ in the \DR scheme is easiest. In this scheme, the stop masses and $X_t$ are, however, renormalisation scale dependent quantities and have no direct relation to physical observables. On the other hand, a renormalisation in the \DR scheme can be advantageous if high-scale SUSY-breaking models are studied. These models impose boundary conditions at some high scale on the \DR parameters. Renormalisation group running is then used to evolve the parameters to the low scale, where physical observables are calculated.


\subsection{Process-dependent OS scheme}

The stop sector can also be renormalised in the OS scheme. This is straightforward for the stop masses, which then correspond to the respective physical masses. To achieve this the counterterms for the diagonal elements of the stop matrix, $m_{\tilde{t}_{1}}^2$ and $m_{\tilde{t}_{2}}^2$, are fixed via the on-shell conditions,
\begin{equation}
  \label{eq:stop_masses_OS_condition}
  \deltaOL m_{\tilde{t}_{1}}^2 = \Re~\Sigma^{(1)}_{\tilde{t}_1 \tilde{t}_1} (m_{\tilde{t}_{1}}^2), \quad
  \deltaOL m_{\tilde{t}_{2}}^2 = \Re~\Sigma^{(1)}_{\tilde{t}_2 \tilde{t}_2} (m_{\tilde{t}_{2}}^2),
\end{equation}
where $\Sigma^{(1)}_{\tilde{t}_i \tilde{t}_i}$ is the one-loop $\tilde{t}_i \tilde{t}_i$ self energy. Note that if we were also including the sbottom sector in this discussion, it would not be possible to renormalise all four stop and sbottom masses on-shell, as a consequence of an $SU(2)_L$ relation --- see e.g.\ the discussion in \ccite{Djouadi:1998sq}.

It is much more difficult to connect the stop mixing parameter to a physical process. An obvious candidate is the decay process $\tilde t_2 \rightarrow \tilde t_1 + h$, which depends on $X_t$ at the tree level. As discussed in \cref{sec:Xt_measurement}, the decay rate and therefore the prospects for experimentally observing this process are, however, highly dependent on the sparticle mass spectrum. In parameter scans this implies that an OS definition of $X_t$ via the process $\tilde t_2 \rightarrow \tilde t_1 + h$ will only be usable for certain parts of the parameter space. The same holds also for processes involving the stop mixing angle at the tree level (see again \cref{sec:Xt_measurement}). Other processes involving a stop--stop--Higgs coupling at the tree level are experimentally difficult to access (e.g.\ $\tilde t_1 \tilde t_2\to h$).


\subsection{Process-independent OS scheme}

Because of the difficulties in defining a process-dependent OS scheme for $X_t$, a process-independent OS scheme is often used in the literature (see e.g.~\ccite{Guasch:1998as,Hollik:2003jj,Heinemeyer:2007aq,Fritzsche:2011nr,Hollik:2014wea}).

The counterterm for the off-diagonal entry of the stop mass matrix is fixed via a symmetric on-shell condition,
\begin{equation}
  \label{eq:mst12_OS_condition}
  \deltaOL m_{\tilde{t}_{12}}^2 = \frac{1}{2} \; \Re \left[\Sigma^{(1)}_{\tilde{t}_1 \tilde{t}_2}(m_{\tilde{t}_1}^2) + \Sigma^{(1)}_{\tilde{t}_1 \tilde{t}_2}(m_{\tilde{t}_2}^2) \right].
\end{equation}
The counterterms $\deltaOL X_t$ or $\deltaOL \theta_{\tilde{t}}$ can then be obtained using the expressions in \cref{eq:stop_CT_relations_a,eq:stop_CT_relations_b,eq:stop_CT_relations_c}.

In MSSM scenarios without \cp-violation in the stop sector, the expression for $\deltaOL \theta_{\tilde{t}}$ reduces to
\begin{equation}\label{eq:dtheta}
  \deltaOL \theta_{\tilde{t}} = \frac{\Re~\Sigma^{(1)}_{\tilde{t}_{1} \tilde{t}_{2}}(m_{\tilde{t}_1}^2) + \Re~\Sigma^{(1)}_{\tilde{t}_{1} \tilde{t}_{2}}(m_{\tilde{t}_2}^2)}{2(m_{\tilde{t}_{1}}^2 - m_{\tilde{t}_{2}}^2)},
\end{equation}
which has been used in \ccite{Yamada:2001px,Brignole:2001jy,Degrassi:2001yf,Heinemeyer:2004xw,Dedes:2003km}.

The disadvantage of such a scheme, where the counterterms are defined in terms of off-diagonal self-energies at a certain value of the squared external momentum, is that the parameters defined in this way are not directly related to physical observables. In fact, the counterterms defined in \cref{eq:mst12_OS_condition,eq:dtheta} will in general be gauge-dependent.


\subsection{\texorpdfstring{\MDR}{MDRbar} and \texorpdfstring{\DR}{DRbar} scheme}
\label{sec:stop_DR_MDR_schemes}

Besides the possibility to connect the parameters to physical 
observables (which is difficult for the stop mixing parameter as discussed above), the OS scheme has a further advantage with respect to the \DR scheme: it ensures the proper decoupling of the gluino. While in the \DR scheme quantum corrections proportional to powers of the gluino mass appear~\cite{Degrassi:2001yf}, these are absent in the OS scheme leaving only a logarithmic dependence on the gluino mass if the gluino mass is much larger than the stop masses. This issue has been discussed in \ccite{Heinemeyer:1998np,Degrassi:2001yf,Muhlleitner:2008yw,Kant:2010tf,Braathen:2016mmb,Aebischer:2017aqa,Kramer:2019fwz,Deppisch:2019iyh,Bahl:2019hmm,Bahl:2019wzx}.

As an alternative to using the OS scheme, the \DR scheme can be modified in order to absorb the corrections enhanced by powers of the gluino mass into the definition of the parameters,
\begin{align}
\left(m_{\tilde t_{L,R}}^\MDR\right)^2(Q) &= \left(m_{\tilde t_{L,R}}^\DR\right)^2(Q)\left[1 + \frac{\alpha_s}{\pi}C_F\frac{|M_3|^2}{m_{\tilde t_{L,R}}^2}\left(1 + \ln\frac{Q^2}{|M_3|^2}\right)\right],\nonumber\\
X_t^\MDR(Q) &= X_t^\DR(Q) - \frac{\alpha_s}{\pi}C_F M_3 \left(1 + \ln\frac{Q^2}{|M_3|^2}\right)\,,
\end{align}
where $\alpha_s\equiv g_3^2/(4\pi)$. As discussed in \ccite{Bahl:2019wzx}, this \MDR scheme consistently avoids the occurrence of terms enhanced by powers of the gluino mass and ensures a proper decoupling behaviour in the limit where the gluino is much heavier than the stops.


\subsection{Mixed schemes}

As an alternative of using a pure OS or a pure \DR/\MDR scheme, the different schemes can also be mixed in the sense that e.g.\ the stop masses are renormalised in the OS scheme while the stop mixing parameter is renormalised in the \DR/\MDR scheme. This particular scheme has the advantage that the stop masses closely correspond to the physical masses while the stop mixing parameter, which is difficult to connect to a physical observable, is renormalised in a simple process-independent scheme.

We stress however that mixed schemes have the disadvantage of the potential occurrence of uncancelled $\epsilon^1$ parts of loop integrals ($\epsilon$ being the UV regulator introduced in dimensional regularisation/reduction). In pure \MS/\DR and OS schemes, all $\epsilon^1$ parts of the involved loop integrals cancel in the final result. For mixed schemes, this is, however, not necessarily the case as noted e.g.\ in \ccite{Borowka:2015ura, Meuserthesis}. This happens for mixed schemes where a quantity is renormalized in the \MS/\DR scheme at the two-loop level but receives a contribution from a one-loop OS counterterm in a sub-loop.

If $\epsilon^1$ parts of loop integrals remain in the final result, the input parameters cannot simply be converted from one scheme to another. For example, a pure \MS/\DR calculation cannot easily be transferred to a calculation in the mixed scheme by a conversion of the input parameter since the \MS/\DR calculation does not contain any $\epsilon^1$ parts of loop integrals and these also cannot be generated by a conversion (without prior knowledge of the structure of the calculations; for examples of a conversion including the $\epsilon^1$ pieces, see e.g.~\ccite{Degrassi:2014pfa, Borowka:2015ura}). Moreover, the occurrence of uncancelled $\epsilon^1$ parts of loop integrals implies that \MS/\DR parameters in a mixed scheme have a different meaning than the corresponding \MS/\DR parameters in a pure \MS/\DR calculation.

Concerning the scheme where the stop mixing parameter 
is renormalised in the \DR or \MDR scheme and the stop masses in the OS scheme, we note that uncancelled  $\epsilon^1$ parts of loop integrals affecting the definition of $X_t$ will only appear for calculations where the stop sector needs to be renormalised at the two-loop level. Only in this case would a one-loop OS counterterm yield a sub-loop contribution to the two-loop \DR/\MDR renormalisation of $X_t$. Thus, in the predictions of the Higgs boson masses this issue of uncancelled $\epsilon^1$ terms will first appear at the three-loop level.

%% file: sec_Xt_Mh.tex
Our discussion in \cref{sec:Xt_measurement} led to the conclusion that confronting the prediction for the mass of the SM-like Higgs boson with the measured experimental value offers the best prospects
for determining the stop mixing parameter. Consequently, an appropriate renormalisation scheme for $X_t$ should be such that a precise Higgs mass prediction can be derived. We start with a discussion of the use of the process-independent OS scheme.

Different methods are employed in precise calculations of the mass of the SM-like Higgs boson. In the fixed-order approach, loop corrections to the inverse Higgs propagator matrix are calculated within the full MSSM. In this framework, a process-independent OS scheme is straightforward to implement. While this approach is suitable for stop masses below or around the TeV scale, its achieved accuracy suffers from the appearance of large logarithmic corrections increasing the size of unknown higher-order corrections for stop masses above the TeV scale. These large logarithmic contributions can be resummed in an EFT approach. In its simplest form, all non-SM particles are integrated out a common mass scale, which is typically set to \msusy. For the calculation of the threshold corrections between the low-energy EFT (e.g.\ the SM) and the high-energy MSSM, typically the limit $v/\msusy\to 0$ is taken, neglecting all higher-order operators appearing in the EFT.\footnote{See also \ccite{Bagnaschi:2017xid} for a departure from this assumption. In this work, the impact of dimension-6 operators to the calculation of the Higgs mass when matching the MSSM onto the SM was considered, and found to be moderate.} For this reason, the EFT calculation is expected to have a lower accuracy for low SUSY scales. Using an OS definition for $X_t$ in the calculation of the threshold corrections would induce large logarithmic terms $\sim \ln \msusy^2/m_t^2$ into the threshold corrections spoiling the underlying assumption of the EFT approach. For this reason, EFT calculations typically employ the \DR scheme (or the \MDR scheme) for the renormalisation of $X_t$.

If $X_t$ could be extracted from a physical observable different from the mass of the SM-like Higgs boson, one would need to extract the input parameter for the EFT calculation of $M_h$ --- $X_t^\DR$ or ($X_t^\MDR$) --- from the physical observable. For simplicity we assume here that this observable is closely related to the process-independent $X_t^\OS$. It is well-known that in the relation between the OS and the \DR definition of $X_t$ large unresummed logarithms can appear in the limit $v/\msusy\to0$. 

The same issue arises in the hybrid calculations combining the fixed-order and the EFT approach in order to obtain a precise $M_h$ prediction for low and high SUSY scales. While the OS scheme can easily be used in the fixed-order part of the calculation, the fixed-order OS quantities have to be converted to \DR (or \MDR) quantities as input for the EFT calculation. This poses the question of whether the logarithms appearing in the relation between $X_t^\OS$ and $X_t^\DR$ can be resummed.

The relation between the OS and the \DR scheme for $X_t$ can be written as follows,
\begin{align}
\label{eq:Xt_conv}
X_t^\OS = X_t^{\DR}(\msusy) \; \frac{m_t^{\DR,\MSSM}(\msusy)}{m_t^\OS} - \frac{1}{m_t^\OS} \deltaOL (m_t X_t) \bigg|_\fin.
\end{align}
We are especially interested in large logarithms appearing in this relation between $X_t^\OS$ and $X_t^\DR$, since such logarithms can potentially spoil the precision of the overall calculation. Both terms on the right-hand side of the \cref{eq:Xt_conv} can contain large logarithms, but these logarithms are of a different origin.

Let us start our discussion with the first term. The relation between the OS and the \DR running top mass can be derived by
\begin{align}
    & m_t^{\DR,\MSSM}(\msusy) = m_t^\OS + \deltaOL m_t^\OS\bigg|_\fin ,
\end{align}
where the finite counterterm has to be evaluated at the renormalisation scale $Q = \msusy$. The finite part of the OS counterterm of the top-quark mass will contain terms like $\ln Q^2/m_t^2$. Since $Q=\msusy$, these terms give rise to large logarithms. The described procedure yields,
\begin{align}
 (\deltaOL m_t^\OS) \Big|_\fin^{Q=\msusy} = m_t^\OS \left[-\left(\frac{\als}{\pi} - \frac{3\alt}{16\pi}\right) \ln \frac{\msusy^2}{m_t^2} + \text{non-log} \right] + \ldots, \label{eq:mt_conv_logs}
\end{align}
where $\alpha_t\equiv y_t^2/(4\pi)=h_t^2/(4\pi s_\beta^2)$ and ``non-log'' is used as a placeholder for terms which do not contain large logarithms, but can contain ``small logarithms,'' i.e.\ logarithms of the ratios $m_A/\msusy,~\vert M_3 \vert/\msusy,~\vert \mu \vert/\msusy$. The ellipsis denotes terms that are not proportional to $\alpha_{s}$ or $\alpha_t$, and which are numerically less important. Expressions containing additional bottom-Yukawa corrections can be found in \cref{app:bottom}. The large logarithms of the form $\ln\msusy^2/m_t^2$ can be resummed by using the top mass defined in the \MS or the \DR scheme at \msusy either in the full MSSM or in the SM.

Now let us proceed with the second term in \cref{eq:Xt_conv}. We will now demonstrate that this term contains large logarithms if the soft-breaking masses of the stops are degenerate,
\begin{align}
\mtL = \mtR = \msusy.
\end{align}
Explicit evaluation of the counterterm $\left.\deltaOL (m_t X_t) \right|_\fin$ in this case shows that it contains the following terms in the limit $\msusy \gg m_t$,
\begin{align}
    \left.\deltaOL (m_t X_t) \right|_\fin ={}& \frac{3\alt}{16 \pi} \; m_t X_t \; \vert \widehat{X}_t \vert^2 \ln \frac{\msusy^2}{m_t^2} + \text{non-log}, \label{eq:Xt_conv_deg_logs}
\end{align}
where $\widehat{X}_{t} = X_{t}/\msusy$. 

\begin{figure}
  \centering
  \includegraphics[scale=1]{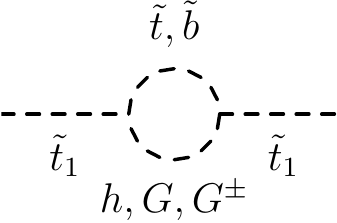}\hspace{.5cm}
  \includegraphics[scale=1]{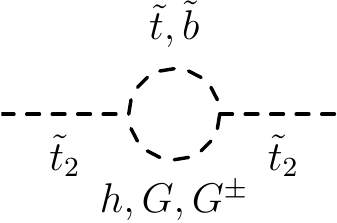}
  \caption{Feynman diagrams generating the large logarithms in \cref{eq:Xt_conv_deg_logs}.}
  \label{fig:Xt_conv_Goldstone_contribution}
\end{figure}

The large logarithm in \cref{eq:Xt_conv_deg_logs} arise from diagrams involving the Goldstone bosons (see \cref{fig:Xt_conv_Goldstone_contribution}).\footnote{The counterterms $\deltaOL m_{\tilde{t}_{12}}^2$ and $\deltaOL m_{\tilde{t}_{21}}^2$ do not give rise to the large logarithms in \cref{eq:Xt_conv_deg_logs}.} Note that these expressions do not depend on the renormalisation scale. Consequently, these large logarithms have a different origin than the logarithms in \cref{eq:mt_conv_logs}.

In \ccite{Bahl:2021rts}, these logarithms have been linked to infrared singularities originating from external-leg corrections. In the present case, the counterterm for the stop mixing angle (see \cref{eq:dtheta}) is of the same form as a $\tilde t_1$--$\tilde t_2$ external leg mixing correction relevant for any process involving an external stop quark. As discussed in detail in \ccite{Bahl:2021rts}, an infrared divergence appears in the limit $m_{\tilde t_1} \rightarrow m_{\tilde t_2}$, which is cured by including Higgs-boson real radiation (note that in the limit $v/\msusy\to0$, the Higgs boson is massless at the tree level). For a finite mass difference, $\Delta m_{\tilde t}^2 = m_{\tilde t_2}^2 - m_{\tilde t_1}^2 = 2 m_t |X_t|$, this infrared limit is manifest in the form of renormalisation scale independent logarithms involving $\Delta m^2$,
\begin{align}
  \ln\frac{\Delta m^2}{\msusy^2} = \ln 2 + \ln \xt - \frac{1}{2}\ln \frac{\msusy^2}{m_t^2},
\end{align}
which are in direct correspondence to the logarithms of \cref{eq:Xt_conv_deg_logs}. These logarithms cannot be resummed by integrating out heavy particles (i.e., the stops in the present case), since they originate from the wave-function normalisation of a heavy particle. Moreover, a dangerous enhancement of the logarithmic terms can also occur due to the trilinear couplings contained in their prefactor, $X_t$ in \cref{eq:Xt_conv_deg_logs}, which can be large. While a resummation can potentially be achieved within the framework of soft-collinear effective field theory, an explicit two-loop calculation in \ccite{Bahl:2021rts} has shown that logarithmic corrections beyond the one-loop order are expected to be relatively small.

The size of unknown higher-order corrections is, however, not the only concern regarding the conversion of $X_t$ from the OS to the \DR scheme. Besides the case of degenerate soft SUSY-breaking masses, of course also the case of non-degenerate soft SUSY-breaking masses needs to be considered. Expanding in powers of $v/\msusy$ while keeping $\mtL \neq \mtR$, we obtain
\begin{align}
    \left.\deltaOL (m_t X_t) \right|_\fin = \frac{\alt}{8 \pi} \; m_t X_t \; \vert \xt \vert^2 \; \left(\frac{2\mtL}{\mtR} \ln \frac{\mtL^2}{\vert \mtL^2 - \mtR^2 \vert} +\frac{\mtR}{\mtL} \ln \frac{\mtR^2}{\vert \mtL^2 - \mtR^2 \vert}  \right) , \label{eq:Xt_conv_nondeg_logs}
\end{align}
where in this case $\xt$ is defined as $\xt = X_t/\sqrt{m_{\tilde{t}_L} m_{\tilde{t}_R}}$. We note that contrary to the expression given in \cref{eq:Xt_conv_deg_logs}, this expression does not contain any large logarithms of the form $\ln\msusy^2/m_t^2$. In the present case, the mass difference $\Delta m_{\tilde t}^2$ regulating the infrared singularity appearing in the wave-function normalisation of the stops is equal to $\mtL^2 - \mtR^2$ --- the additional term $2 m_t |X_t|$ in the stop mass difference $\Delta m_{\tilde{t}}^2$ can be neglected here in the limit $v/\msusy \to 0$. 

To summarise, the conversion formula for the stop mixing parameter $X_t$ in the heavy SUSY limit  --- $v/\msusy\to 0$ --- for the case of the scenario with degenerate squark soft-breaking masses reads,
\begin{align}
  X_t^{\DR}(\msusy)\bigg|_{\mtL=\mtR} &= X_t^\OS\left\{1 + \left[\frac{\als}{\pi} - \frac{3\alt}{16\pi}\left(1 - \frac{\vert X_t \vert^2}{\msusy^2}\right)\right]\ln\frac{\msusy^2}{m_t^2}\right\} + \text{non-log}.
\end{align}
For the case where the stop soft-breaking masses are non-degenerate, the corresponding formula takes the form,
\begin{align}
  X_t^{\DR}(\msusy)\bigg|_{\mtL\neq \mtR} = X_t^\OS\left\{1 + \left[\frac{\als}{\pi} - \frac{3\alt}{16\pi}\right]\ln\frac{\msusy^2}{m_t^2}\right\} + \text{non-log}.
\end{align}
This clearly shows that in the limit $v/\msusy \to 0$ no smooth transition between the cases $\mtR = \mtL$ and $\mtR \neq \mtL$ exists.

\begin{figure}
  \centering
  \includegraphics[width=0.9\textwidth]{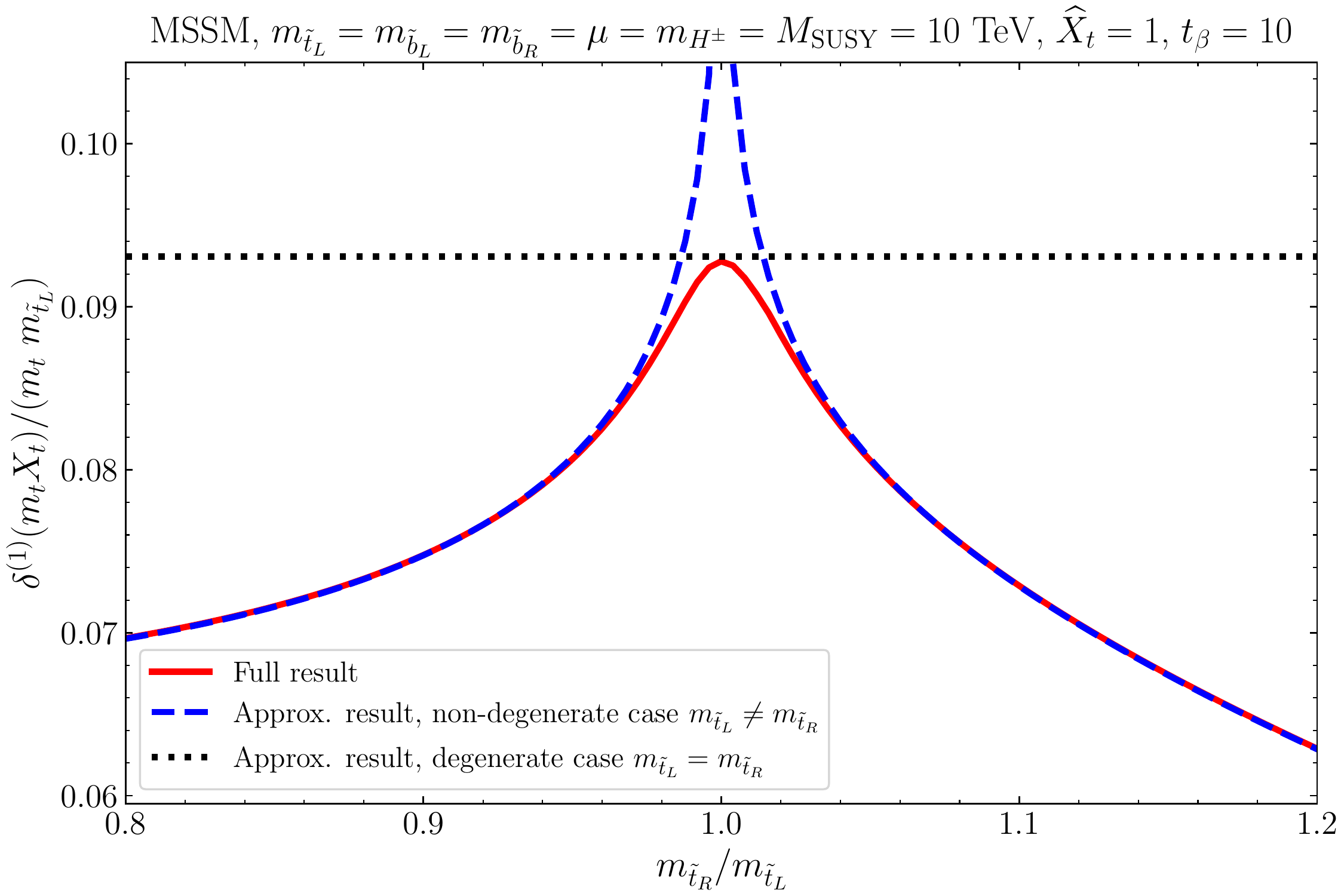}
  \caption{Numerical comparison of $\deltaOL (m_t X_t)/ (m_t \mtL)$ calculated without any expansion (red solid), calculated in the limit $v/\msusy\to 0$ and $\mtL\neq\mtR$ (blue dashed), as well as calculated in the limit $v/\msusy\to0$ and $\mtL = \mtR$ (black dotted).}
  \label{fig:dXtMT_deg_vs_nondeg}
\end{figure}

The formula that is not expanded in the limit $v/\msusy\to 0$ does not show this behaviour. To illustrate this, we consider a MSSM scenario in which $\mu = 10 \tev$, all soft-breaking masses except for $\mtR$ are equal to $\msusy =  10 \tev$, the stop mixing parameter equals $X_t = \msusy$, and $\tan \beta = 10$. In \cref{fig:dXtMT_deg_vs_nondeg}, we show the behaviour of the counterterm $\deltaOL (m_t X_t)$, normalised by the top mass $m_t$ and the soft mass $\mtL$, neglecting corrections proportional to the bottom Yukawa and the strong gauge coupling for simplicity. 

The solid red curve corresponds to the full expression of the order $\order{\alt}$ contribution to $\deltaOL (m_t X_t)/ (m_t \mtL)$. The horizontal black dotted line indicates the value corresponding to the approximate expression in \cref{eq:Xt_conv_deg_logs}, while the blue dashed curve shows the behaviour of \cref{eq:Xt_conv_nondeg_logs}. Since the SUSY scale is much heavier than the EW scale, the blue dashed curve yields a good approximation of the behaviour of the red curve for $\left \vert \mtR/\mtL - 1 \right \vert \gtrsim 0.05$. On the other hand it starts to deviate from the red one for values of $\mtR/\mtL$ close to one, and for $\mtR/\mtL = 1$ its value is not defined. At this point the full expression is well approximated by the result that is indicated by the black dotted line.

The one-loop contribution to the counterterm $\deltaOL (m_t X_t)$ proportional to the strong gauge coupling does not exhibit the behaviour described above. More specifically, there is no large logarithm emerging in the $\order{\als}$ expression for $\left.\deltaOL (m_t X_t) \right|_\fin$ in the heavy SUSY limit regardless whether the squark soft-breaking masses are degenerate or non-degenerate. Furthermore, for the corrections of order $\order{\als}$ a smooth transition between the two mentioned scenarios exists (since the particles appearing in the loop are fermions or gauge bosons and not scalars).

While it is possible to use the full unexpanded $X_t^\OS \to X_t^\DR$ conversion, this would mix different orders in the EFT expansion. For this reason and since $M_h$ is the most promising observable to determine $X_t$ (as discussed in \cref{sec:Xt_measurement}), it seems preferable to use a \DR renormalisation of $X_t$ in the fixed-order calculation as well. If an extraction of $X_t$ from another observable than $M_h$ is found to be possible in the future, the issues of large logarithmic contributions to the extraction will, however, reappear. Furthermore, as discussed above, a full \DR scheme is problematic because of non-decoupling effects for the case where the gluino is much heavier than the stops, while a mixed scheme where $X_t$ is renormalised in the \DR scheme and the stop masses are renormalised in the OS scheme can lead to complications at higher orders because of uncancelled terms of $\mathcal{O}(\epsilon^1)$ arising from the loop integrals. A possible solution is to start with a \DR renormalisation, and then reparametrise the quantities in the calculation without $\mathcal{O}(\epsilon^1)$ pieces. This, however, would not be convenient for a calculation initially based on the OS scheme. 

While employing an \MDR scheme for $X_t$ allows avoiding the first issue of unphysical non-decoupling effects, the problem of mixed schemes persists.  Thus, while at present it seems difficult to define a ``best'' scheme for $X_t$ that would be suitable also for future higher-order evaluations, a mixed scheme where a \MDR renormalisation of $X_t$ is combined with an OS renormalisation of the stop masses appears to be the preferred choice in view of the currently available level of higher-order corrections.

%% file: sec_conclusions.tex
An interesting possible feature of extensions of the SM with additional scalars is the existence of new types of interactions, like mass-dimensionful trilinear couplings, which are not induced by a vacuum expectation value. In this article, we have focused on the case of the MSSM, in which such a trilinear coupling controls the interaction between the scalar top quarks and the Higgs bosons, in the form of the stop mixing parameter $X_t$. In the event of a discovery of a BSM scalar sector, measuring the interactions between the different states will be of paramount importance to properly characterise the underlying model. We have discussed several approaches to access $X_t$ via experimental measurements. 

As a starting point, we illustrated the fact that the knowledge of just the two stop masses is not sufficient to determine $X_t$ since additional information or some assumptions about the stop soft SUSY-breaking masses would be required in this case.
While the measurement of the stop mixing angle in combination with measurements of the two stop masses can in principle be used to determine $X_t$ together with the two stop soft SUSY-breaking masses, a precise determination of the stop mixing parameter is only possible in this way if the mass scale of the two stops is close to the TeV scale. The sensitivity to $X_t$ rapidly deteriorates as the SUSY breaking scale increases. It should be stressed that this finding is valid even for the case of a collider of sufficient energy that has the capability for precise measurements of the stop masses and the stop mixing angle. The reason for the loss of sensitivity for higher values of \msusy are contributions that scale like $m_t/\msusy$. Another option that we investigated was extracting $X_t$ from the measurement of a decay process like $\tilde{t}_2\to \tilde{t}_1 + h$, but this has the disadvantage that the decay width, or its branching ratio, can be significantly suppressed if the stops are approximately mass-degenerate or in scenarios allowing other decays of stop quarks (e.g.\ if light electroweakinos are present). 

As a result of our investigations, we found that the observable offering the best prospects for accessing $X_t$ appears to be the mass of the SM-like Higgs boson, $M_h$. The parameter $X_t$ enters the prediction for $M_h$ starting from the one-loop level. We showed that with the input of the stop masses and of $\tan\beta$, the value of $X_t$ can be determined from the measured value of $M_h$ to a high level of accuracy. A further remarkable feature of the prediction for the mass of the SM-like Higgs boson in this context is that it retains a sizeable dependence on $X_t$ even for SUSY scales as high as 100 TeV. Moreover, moderate variations of the SUSY breaking parameters do not alter the prediction of $M_h$ very significantly. Thus, even if the knowledge of the SUSY spectrum is only quite limited, this is not expected to spoil the determination of $X_t$ from $M_h$. 

Next, we compared different possible choices of renormalisation schemes for the stop sector. The simplest option is to renormalise all stop-sector quantities, i.e.\ the stop masses and the mixing parameter, in the \DR scheme. In addition to its simplicity, this choice can also be useful when investigating scenarios with high-scale boundary conditions employing the renormalisation-scale running of parameters. However, the \DR scheme can also give rise to the known issue of unphysical non-decoupling effects  --- which for instance occur if there is a large hierarchy between the gluino and stop masses. One way to avoid such effects is to adopt an on-shell renormalisation scheme. While for the stop masses an OS renormalisation implies a unique definition of the counterterm, for $X_t$ several choices are possible, depending on whether one relates the $X_t$ counterterm to the calculation of a physical process, like e.g.\ $\tilde t_2\to\tilde t_1+h$, or whether one instead relates this counterterm to the counterterms of the stop mass matrix. Another choice of renormalisation scheme that allows avoiding unphysical enhancements is the \MDR scheme, in which the finite parts of the stop-mass and $X_t$ counterterms are defined in such a way as to absorb the contributions involving powers of the gluino mass. We furthermore considered the possibility of adopting a mixed renormalisation scheme, i.e.\ renormalising some parameters on-shell but keeping others in the \DR scheme. Such a choice, however, gives rise to the potential issue of a non-cancellation of $\epsilon^1$ parts of loop integrals (from the three-loop level onwards), which changes the physical meaning of \DR parameters compared to a pure \DR scheme and prevents direct scheme conversions.

In the last part of our work, we considered the renormalisation of $X_t$ in the specific context of the prediction for the mass of the SM-like Higgs boson. As we have demonstrated above that $M_h$ is the most promising observable to determine $X_t$, it is crucial to assess to which extent the choices of the renormalisation prescription for $X_t$ are compatible with the different approaches for computing $M_h$ -- fixed order, EFT, and hybrid. In a fixed-order calculation of $M_h$, adopting an OS renormalisation for $X_t$ is a simple choice, which is straightforward to implement. On the other hand, when performing an EFT calculation of the Higgs mass, it is preferable to renormalise $X_t$ in the \DR or \MDR scheme. This is also the case for hybrid computations, for which it is best to use a \DR/\MDR scheme for the EFT part. This raises the issue of the conversion between OS and \DR/\MDR schemes for $X_t$. We have pointed out the existence of possible large logarithmic terms in this conversion in the limit $\msusy\gg v$, which can cause a loss of accuracy of the entire $M_h$ calculation. We have clarified the source of different types of large logarithmic terms and investigated the possibility of resumming them. A first type of large logarithms stems from the contribution involving the OS-scheme counterterm for the top-quark mass evaluated at $Q=\msusy$. These logarithms can be resummed via renormalisation-group running. However, a second, more problematic, type of terms arises from the finite part of the counterterm of the off-diagonal stop mass matrix when expanded in the $\msusy\gg v$ limit. These terms cannot be resummed with renormalisation group methods (a resummation within the framework of a soft-collinear effective theory should on the other hand in principle work), and moreover, their form depends on whether one performs the expansion in powers of $v/\msusy$ for the case of degenerate or non-degenerate soft stop masses. We found that there exists no smooth transition between these two cases. Avoiding the expansion in $v/\msusy$ is also not a viable option as this would mix orders of the EFT expansion in the $M_h$ calculation. We, therefore, conclude that for the determination of the stop mixing parameter $X_t$ from confronting a hybrid calculation of $M_h$ with the measured value, the most advantageous choice is to adopt a \DR/\MDR renormalisation for $X_t$. We note, however, that starting at the three-loop level this scheme will be affected by the problem related to uncancelled $\epsilon^1$ parts of loop integrals described above.

Finally, we want to remark again that mass-dimensionful trilinear couplings do not only appear in the MSSM but also in other BSM models. While the present study is focused on the stop mixing parameter in the MSSM, we expect that many of the difficulties identified in this study related to measuring and renormalising trilinear couplings will also appear in other models. Therefore, the situation can be expected to be more problematic in non-supersymmetric models, where the Higgs mass cannot be used to constrain trilinear couplings. As we have shown, this is the case even if the new particles can be produced at the LHC or a future collider, and the corresponding mixing angle can be measured.

%% file: app_bottom.tex
In this Appendix, we present expressions for the conversion of $X_t$ between the OS and the \DR scheme including corrections controlled by the bottom Yukawa coupling.

In the case of $\mtL = \mtR = \mbL = \mbR = \msusy$, the finite part of the $\deltaOL (m_t X_t)$ counterterm reads 
\begin{align}
    \left.\deltaOL (m_t X_t) \right|_\text{fin} ={}&  \frac{\alt}{4 \pi} \; m_t X_t \; \vert \widehat{X}_t \vert^2 \; \ln \frac{\msusy^2}{2 m_t \vert X_t \vert} \nonumber\\
    & + \frac{e^{i \pXt} \msusy}{64 \pi^2 v^2} \left(m_t \vert \xt \vert - m_b \vert \xb \vert \right)^3 \; \ln \frac{\msusy^2}{\left \vert m_t \vert X_t \vert - m_b \vert X_b \vert \right \vert} \nonumber\\
    & + \frac{e^{i \pXt} \msusy}{64 \pi^2 v^2} \left(m_t \vert \xt \vert + m_b \vert \xb \vert \right)^3 \; \ln \frac{\msusy^2}{\left \vert m_t \vert X_t \vert + m_b \vert X_b \vert \right \vert} \nonumber\\
    & + \text{non-log} = \nonumber\\
    ={}& \frac{3\alt}{16 \pi} \; m_t X_t \; \vert \widehat{X}_t \vert^2 \ln \frac{\msusy^2}{M_t^2} + \frac{3\alb}{16 \pi} \; m_t X_t \; \vert \widehat{X}_b \vert^2 \ln \frac{\msusy^2}{M_t^2}\nonumber\\
    & + \text{non-log},
\end{align}
in the limit $v/\msusy \to 0$. $\alpha_b \equiv y_b^2/(4\pi)$ with $y_b$ being the bottom-Yukawa coupling.

For $\mtL \neq \mtR, \mbL \neq \mbR, \mtR \neq \mbR$, we obtain
\begin{align}
    \deltaOL &(m_t X_t) \bigg|_\text{fin} =\nonumber\\
    ={}& \frac{\alt}{8 \pi} \; m_t X_t \; \vert \xt \vert^2 \; \left(\frac{2\mtL}{\mtR} \ln \frac{\mtL^2}{\vert \mtL^2 - \mtR^2 \vert} +\frac{\mtR}{\mtL} \ln \frac{\mtR^2}{\vert \mtL^2 - \mtR^2 \vert}  \right) \nonumber\\
    &  + \frac{\alb}{8\pi} \; m_t  X_t \; \vert \xb \vert^2 \left( -\frac{\mbR (\mbR^2 - 2 \mtL^2 + \mtR^2)}{\mtL (\mtL^2 - \mtR^2)} \ln \frac{\mbR^2}{\vert \mbR^2 - \mtL^2\vert} \right. \nonumber\\ 
    & \hspace{3.4cm}+ \frac{\mbR \mtL (\mbR^2 - \mtR^2)^2}{(\mbR^2 - \mtL^2) \mtR^2 (\mtL^2 - \mtR^2)} \ln \frac{\mbR^2}{\vert \mbR^2 - \mtR^2 \vert} \nonumber\\
    & \hspace{3.4cm}\left. - \frac{\mbR \mtL (\mtL^2 - \mtR^2)}{\mtR^2 (\mbR^2 - \mtL^2)}
    \ln \frac{m_{\tilde{t}_L}^2}{\vert m_{\tilde{t}_L}^2 - m_{\tilde{t}_R}^2 \vert} + \frac{2 \mbR \mtL}{\mbR^2 - \mtL^2} \ln \frac{\mbR^2}{\mtL^2} \right)  \nonumber\\
    & + \ldots ,
\end{align}
where $\xb = X_b/\sqrt{m_{\tilde{b}_L} m_{\tilde{b}_R}}$. The ellipsis denotes further non-logarithmic terms. This means that the expression does not contain any terms $\propto \ln\msusy^2/m_t^2$.

For $\mbL = \mbR, \; \mtL \neq \mtR$, we obtain
\begin{align}
\left.\deltaOL (m_t X_t) \right|_\text{fin} = \frac{\alb}{16 \pi} \; m_t X_t \; \frac{\vert X_b \vert^2}{\mbL^2} \ln \frac{\msusy^2}{M_t^2} + \text{non-log},
\end{align}
where here the non-logarithmic terms include logarithms not including a light SM mass (e.g.\ $\ln\mtL^2/|\mtL^2-\mtR^2|$).

For $\mtL = \mtR, \; \mbL \neq \mbR$, the large logarithms in the conversion formula take the following form,
\begin{align}
\left.\deltaOL (m_t X_t) \right|_\text{fin} \supset \frac{3\alt}{16 \pi} \; m_t X_t \; \frac{\vert X_t \vert^2}{\mtL^2} \ln \frac{\msusy^2}{M_t^2} + \text{non-log}.
\end{align}

To summarise, the conversion formula for the stop mixing parameter $X_t$ in the heavy SUSY limit for the case of the scenarios with fully degenerate squark soft-breaking masses reads,
\begin{align}
  X_t^{\DR}(\msusy) &= X_t^\OS\left\{1 + \left[\frac{\als}{\pi} - \frac{3\alt}{16\pi}\left(1 - \frac{\vert X_t \vert^2}{\msusy^2}\right) \right.\right.\nonumber\\
   &\left.\left.\hspace{3.15cm} + \frac{3\alb}{16\pi}\left(1 + \frac{\vert X_b \vert^2}{\msusy^2}\right)\right] \ln\frac{\msusy^2}{M_t^2}\right\} + \text{non-log}.
\end{align}
For the case where the stop soft-breaking masses are degenerate, but the sbottom soft-breaking masses are non-degenerate, this formula takes the form,
\begin{align}
  X_t^{\DR}(\msusy) ={}& X_t^\OS\left\{1 + \left[\frac{\als}{\pi} - \frac{3\alt}{16\pi}\left(1 - \frac{\vert X_t \vert^2}{\mtL^2}\right) + \frac{3\alb}{16\pi}\right] \ln\frac{\msusy^2}{M_t^2}\right\} \nonumber\\
  &+ \text{non-log}.
\end{align}
If instead the sbottom soft-breaking masses are equal to each other but the stop soft-breaking masses are non-degenerate, the logarithmic terms in the conversion formula read,
\begin{align}
  X_t^{\DR}(\msusy) ={}& X_t^\OS\left\{1 + \left[\frac{\als}{\pi} - \frac{3\alt}{16\pi} + \frac{3\alb}{16\pi}\left(1 + \frac{\vert X_b \vert^2}{3\mbR^2}\right)\right] \ln\frac{\msusy^2}{M_t^2}\right\} \nonumber\\
  & + \text{non-log}.
\end{align}
In all other cases the conversion formula reads,
\begin{align}
X_t^{\DR}(\msusy) = X_t^\OS \left\{1 + \left[\frac{\als}{\pi} - \frac{3\alt}{16\pi} + \frac{3\alb}{16\pi} \right]\ln\frac{\msusy^2}{M_t^2} \right\} + \text{non-log}.
\end{align}